\documentclass{stavanger}

\usepackage[utf8]{inputenc}
\usepackage{amssymb,amsmath,amsfonts,amsthm}
\usepackage{hyperref}
\usepackage{natbib}
\usepackage{graphicx}
\usepackage{tikz}
\usepackage{cleveref}
\usepackage{ulem} 
\usepackage{placeins}

\startlocaldefs

\endlocaldefs

\begin{document}


\begin{frontmatter}

\begin{fmbox}


\dochead{Research Article - Preprint}{FP}


\title{Proper static potential in classical lattice gauge theory at finite T}


\author[
   addressref={aff1,aff2},                   	  
   email={alexander.lehmann@uis.no}   		  
]{\inits{AL}\fnm{Alexander} \snm{Lehmann}}
\author[
   addressref={aff2},                   	  
   corref={aff2},                     		  
   email={alexander.rothkopf@uis.no}   		  
]{\inits{AR}\fnm{Alexander} \snm{Rothkopf}}


\address[id=aff1]{
  \orgname{Institute for Theoretical Physics, Heidelberg University}, 	 
  \street{Philosophenweg 16},                     		 
  \postcode{69120}                               			 
  \city{Heidelberg},                              				 
  \cny{Germany}                                   				 
}
\address[id=aff2]{
  \orgname{Faculty of Science and Technology}, 	 
  \street{University of Stavanger},                     		 
  \postcode{4021}                               			 
  \city{Stavanger},                              				 
  \cny{Norway}                                   				 
}



\end{fmbox}


\begin{abstractbox}

\begin{abstract} 
We compute the proper real-time interaction potential between a static quark and antiquark in classical lattice gauge theory at finite temperature. Our central result is the determination of the screened real-part of this potential, and we reconfirm the presence of an imaginary part. The real part is intimately related to the back-reaction of the static sources onto the gauge fields, incorporated via Gauss's law. Differences in the treatment of static sources in quantum and classical lattice gauge theory are discussed.
\end{abstract}



\end{abstractbox}


\end{frontmatter}


\section{Introduction}
\label{sec:introduction}

The bound states of heavy quarks and antiquarks, so called heavy quarkonium, constitute a unique laboratory to scrutinize the physics of the strong interactions (for a broad overview see Ref.~\cite{Brambilla:2010cs}). In particular, studying their production in relativistic heavy-ion collisions promises vital insight into the properties of quarks and gluons under the extreme conditions of high temperature and density which are present shortly after the Big bang (for a recent review see Ref.~\cite{Rothkopf:2019ipj}). In contrast to the earliest moments of the universe, a heavy-ion collision produces energetic debris that are not in static thermal equilibrium (for selected perspectives see, e.g., Refs.~\cite{Strickland:2020hyg,Berges:2020fwq,Shen:2020gef,Akamatsu:2020lej,Kurkela:2019kdu} and references therein). At the time of incident, hard partonic scatterings may convert the vast amounts of kinetic energy of the incoming projectiles into particles with high velocity or large masses (compared e.g. to the characteristic scales of QCD $\Lambda_{\rm QCD}/p_Q  \ll 1$). It is here where the constituents of quarkonium particles are born. On the other hand, the lighter particles, which are created for example from the fragmentation of the strong initial glasma color fields, quickly thermalize locally and form a liquid-like and expanding quark-gluon plasma. One challenge in this field of study is to understand how the hard particles, which are produced in the initial stages, propagate in the presence of the quasi-thermal hot environment formed by the light degrees of freedom. 

Quarkonium in equilibrium with its surroundings has been studied thoroughly in the past using lattice QCD simulations and effective field theories (for some recent works see \cite{Larsen:2019zqv,Offler:2019eij,Larsen:2019bwy,Kim:2018yhk,Burnier:2017bod}), potential models (see e.g. \cite{Burnier:2015tda,Burnier:2016kqm}), QCD sum rules \cite{Song:2020kka} and holography (see e.g.  \cite{Braga:2017bml,Grigoryan:2010pj,Fujita:2009ca}). In recent years, thanks to advances in real-time methods, the focus has shifted into the realm of non-equilibrium physics, heavily motivated by the phenomenologically relevant setting of relativistic heavy-ion collision. A fruitful exchange of ideas between the high-energy nuclear physics community and the condensed matter physics community (see e.g. \cite{Borghini:2011ms,Akamatsu:2011se,Akamatsu:2012vt}) has given momentum to the ongoing development of an open quantum systems description of heavy quarkonium (for the most recent works see e.g. \cite{Brambilla:2020qwo,deJong:2020tvx,Yao:2020eqy,Yao:2020xzw,Alund:2020ctu,Miura:2019ssi,Brambilla:2019tpt,Yao:2018sgn,Blaizot:2018oev,Brambilla:2017zei,Kajimoto:2017rel,Yao:2017fuc}) in contact but not necessarily in equilibrium with its hot environment (for a recent review see \cite{Akamatsu:2020ypb}). A central ingredient in the development of these real-time descriptions is the inherent separation of scales between the heavy-quark rest mass and the other characteristic scales, such as energy density and the QCD scale $\Lambda_{\rm QCD}$. In vacuum, this separation of scales has helped to establish that the physics of heavy quarkonium bound states may be captured reliably by a non-relativistic potential description. Formalized in the effective field theories Non-relativistic QCD (NRQCD) and potential NRQCD  (see \cite{Brambilla:2004jw} for a review), it has been established, at least in a weakly-coupled context, to what extent such a potential picture is also applicable at finite temperature \cite{Laine:2006ns,Beraudo:2007ky,Brambilla:2008cx}. The determination of the interaction potential non-perturbatively using lattice QCD simulations is an active field of ongoing research \cite{Rothkopf:2011db, Burnier:2014ssa, Burnier:2016mxc, Petreczky:2017aiz, Petreczky:2018xuh, Lafferty:2019jpr}. Recent work on quarkonium as open-quantum system has shown how the real- and imaginary part of this in general complex valued static heavy quark potential govern the evolution of quarkonium states in a hot medium \cite{Kajimoto:2017rel, Miura:2019ssi}. While the in-medium real part informs us about the screening of the interaction between the heavy quark-antiquark-pair, the imaginary part encodes how scattering with gluons of the surrounding medium over time leads to color decoherence.

The complex static interquark potential at finite temperature thus plays a vital role in the real-time description of quarkonium bound states. It is now understood that we may access its values non-perturbatively, by inspecting the rectangular real-time Wilson loop
\begin{align}
    W_\square(t,r) = \left\langle {\rm Tr}\Big[{\cal P}{\rm exp}\Big[ ig\int_{{\cal C}_{\square(r,t)}} dx^\mu A^\mu(x)\Big]\Big]\right\rangle\,,\label{eq:wilsonloop}
\end{align}
which resides on the path that a pair of static color sources traces out as it evolves in real-time. $A_\mu$ refers to the gauge field of the strong interactions, and, as a correlation function, the Wilson loop is evaluated in path-ordered fashion, indicated by the operator ${\cal P}$. Its gauge invariance follows from taking the color trace, denoted above by ${\rm Tr}$. As was shown in detail in \cite{Burnier:2012az}, if the time evolution of a pair of color sources, described by $W_\square(t,r)$, proceeds according to a static Schr\"odinger equation, we may use it to define the corresponding potential via
\begin{align}
    V(r)=\lim_{t\to\infty}  i\partial_t W_\square(t,r)/  W_\square(t,r).  \label{eq:defpot}
\end{align}
In the context of the present paper, we consider times bigger than the time corresponding to multiple gluon mediated exchanges as late times. These gluon mediated exchanges may at late times then be collectively replaced by an instantaneous interaction potential. In practice, evaluating \cref{eq:defpot} often presents technical difficulties. In turn it has been worked out (see e.g. Ref.~\cite{Burnier:2013fca}) that the values of the potential may be reliably and efficiently extracted from the spectral function of the Wilson loop instead. The spectral function is the unique real-valued function describing the different incarnations of quantum field theoretical correlation functions such as of time ordered, retarded, or Matsubara type. If a time independent potential emerges at late times according to \eqref{eq:defpot}, one can show that there must exist a lowest lying peak structure of skewed Breit-Wigner form in the spectral function, whose position and width encode the real and imaginary part of the potential respectively. Around the peak maximum, we may describe it with the following functional form
\begin{align}
   \begin{split}
     \rho_\square(r,\omega)=&\frac{e^{{\rm Im}[\sigma_\infty](r)}}{\pi}\cdot\\&\frac{|{\rm Im}[V](r)|{\rm cos}[{\rm Re}[\sigma_\infty](r)]-({\rm Re}[V](r)-\omega){\rm sin}[{\rm Re}[\sigma_\infty](r)]}{ {\rm Im}[V](r)^2+ ({\rm Re}[V](r)-\omega)^2}\\
    &+c_0(r)+c_1(r)t_{Q\bar Q}({\rm Re}[V](r)-\omega)\\
    &+c_2(r)t_{Q\bar Q}^2({\rm Re}[V](r)-\omega)^2+\cdots.
    \end{split}
    \label{eq:sBWfit}
\end{align}
The values of the static potential have been investigated in the past through conventional lattice QCD simulations. However, the fact, that these simulations are carried out in Euclidean time, necessitates the solution of an ill-posed inverse problem to extract the spectral function, often attacked using methods of Bayesian inference (see e.g. \cite{Rothkopf:2019dzu}). While robust estimates of the real part of the potential have been obtained in this fashion, access to the imaginary part is still severely limited. The reason is that the determination of a spectral width requires significantly higher input data quality than that of spectral peak positions.

In this study, we set out to investigate the complex interquark potential in a genuine real-time setting. Since the notorious sign problem of the quantum path integral prohibits its direct numerical evaluation in Minkowski-time, we will have to agree to a compromise. In case of this study, it amounts to neglecting the quantum fluctuations in the theory and simply focusing on the statistical fluctuations, wherein such statistical fluctuations may be introduced through a thermal medium. I.e., we will resort to the classical statistical approximation of Yang-Mills theory to investigate the binding properties of a pair of static color charges at finite temperature. The classical statistical treatment of gauge fields has a long history in the context of research on Baryogenesis (see e.g. Ref.~\cite{DOnofrio:2014rug} and references therein), reheating in the early universe in Refs.~ \cite{Micha:2004bv,Berges:2008wm,Berges:2016nru,Chatrchyan:2020syc} and more recently in the study of the overoccupied glasma in the early stages of relativistic heavy-ion collisions ( for a review see Ref.~\cite{Berges:2020fwq}). The static interquark potential has been studied in the classical statistical approximation first in Ref. \cite{Laine:2007qy}. Here, we will extend and improve on those results, based on work presented in A.~Lehmann's doctoral thesis \cite{phdthesis_AlexanderLehmann}.

While the non-equilibrium properties of several phenomenologically relevant systems can be captured quantitatively by classical statistical simulations, we can only expect to gain qualitative insight in the case of thermal equilibrium. 
The reason is the classical Raleigh Jeans divergence. Thermal modes saturate the spectrum of any classical statistical computation and in turn the effects of, e.g., charge screening, dominated by modes around the cutoff, become dependent on this UV cutoff.  Viewed through the lens of lattice perturbation theory, one furthermore predicts \cite{Arnold_1997,WATSON_1939,Glasser_2000} that the Debye mass changes with the square root of the temperature at fixed lattice spacing
\begin{align}
(m_D^{\rm lPT})^2 = 2 g^2TN_c \frac{\Sigma_{\rm lat}}{4\pi a_s}, \quad \Sigma_{\rm lat}=\Gamma^2(\frac{1}{24})\Gamma^2(\frac{11}{24})\frac{\sqrt{3}-1}{48\pi^2}, \label{eq:DebyeMassLPT}
\end{align}
 which is in contrast to the full quantum theory, where the Debye mass is proportional to the temperature itself.

The absence of a well-defined continuum limit precludes us, e.g., from attaching physical units to the simulations. There exist prescriptions of how to make possible a continuum limit by amending the classical simulation with a perturbative treatment of modes close to and above the lattice cutoff (see e.g. \cite{Bodeker:1998hm,Bodeker:1999gx}). For our purposes of qualitatively identifying the form of the interaction potential, the naive implementation suffices, however. Keeping these caveats in mind, classical statistical simulations, as one of the very few direct real-time methods available, promise vital insight into the dynamics of binding of color sources in Yang-Mills theory and provide guidance to understand the fully quantum theory so far only accessible through the lens of static Euclidean lattice QCD simulations.

In the following \cref{sec:statsource}, we discuss the differences in treatment of static sources between conventional Euclidean lattice QCD simulations and those carried out in the classical statistical approximation. \Cref{sec:nummethods} briefly describes the numerical methods used in our study before we present our findings in \cref{sec:numres}. The paper closes with a conclusion and outlook in \cref{sec:conclusion}.

\section{Static Sources in Lattice Gauge Theory}
\label{sec:statsource}
The study of the binding properties of static charges in the presence of a medium of light charge carriers goes back to the works of Debye and H\"uckel \cite{huckel1923theory}. From general thermodynamic considerations and use of linear response theory, they deduced classically that in such a system the interactions between the static charges will be screened, i.e., the long-ranged Coulomb interaction will become short ranged with a characteristic screening radius given by the temperature dependent density of light charge carriers. The study of the real-part of the interquark potential in conventional lattice QCD supports this conclusion also for the strong interactions, showing clear signs for a bound state sustaining real-part of the potential in vacuum and a gradual weakening of its values towards a screened form at high temperatures \cite{Burnier:2014ssa}.

What would be the intuitive expectation for a classical theory of strong interactions? In the $\hbar\to0$ limit, Yang-Mills theory may be seen as a non-linear extension of Maxwell's electrodynamics. Since charges interact in electrostatics, there is no a priori reason to suspect otherwise in the non-linear theory. Irrespective of the sign of the interaction, we would expect that its strength is encoded in the finite real-part of a classical interaction potential.

The results of Ref. \cite{Laine:2007qy} hence are puzzling. The authors set out to study the static potential in classical statistical lattice gauge theory based on the real-time Wilson loop introduced in \cref{sec:introduction}. They observe that its values are purely real. According to \cref{eq:defpot}, if the Wilson loop goes over into a single exponential behavior at late times it depends on the values of the potential as
\begin{align}
   \lim_{t\to\infty} W_\square(r,t)\sim {\rm exp}\big[ -i t V(r) \big]
\end{align}
A purely real $ W_\square(r,t)$ thus corresponds to a purely imaginary $V(r)$. 

Does this mean that the real-part is screened to such small values that it is practically invisible? Or does it mean that there are no interactions between color sources present at all in the classical statistical theory? In this study, we propose and show that the potential between static color sources is complex. It possesses a non-vanishing real and imaginary part. The previously observed absence of a real part originates from subtle differences in how static sources have to be treated in the quantum versus the classical theory.

In fully quantum lattice gauge theory, the physics of static color sources are conventionally investigated by computing the expectation values of the imaginary time Wilson loop $W_\square(-i\tau,r)$ in the Euclidean path integral. At first, simulations of the gauge (and possibly also light fermion) degrees of freedom are carried out, which are oblivious of the physics of the static sources that we wish to investigate. It is the evaluation of the Wilson loop itself that introduces these charges into the system. How is this accomplished? To not encumber us with technical difficulties, let us assume that we are in spatial axial gauge, so that only the temporal stretches of the Wilson loop remain relevant:
\begin{align}
   &W_\square(\tau,r=|{\bf x}_2-{\bf x}_2|) \\
   \nonumber&= \frac{1}{Z_{\rm no\,src}}\int {\cal D}[A] {\rm Tr}\Big[ {\rm exp}\Big[ig\int dt A_0({\bf x}_1,t)\Big]{\rm exp}\Big[-ig\int dt A_0({\bf x}_2,t)\Big] \Big] e^{ -S_{\rm gluon}[A]}\\
  \nonumber&= \frac{1}{Z_{\rm no\,src}}\int {\cal D}[A] {\rm exp}\Big[ -S_{\rm gluon}[A]  - g\int d^4x\,  {\rm ReTr}\big[ A_0(x) M \big( \delta^{(3)}({\bf x}-{\bf x}_1)-\delta^{(3)}({\bf x}-{\bf x}_2) \big) \big]\Big]\\
  \nonumber &= \frac{1}{Z_{\rm no\,src}}\int {\cal D}[A]
  {\rm exp}\Big[ -S_{\rm gluon}[A]  - g\int d^4x\,  {\rm ReTr}\big[ A_0(x)j_0(x,{\bf x}_1,{\bf x}_2) \big]\Big]\\
\nonumber & = \frac{Z_{\rm src}(\tau,r)}{Z_{\rm no\,src}}.
\end{align}
The matrix $M$ denotes a $3\times3$ matrix that arises when absorbing the two exponentials into the exponent of the Feynman weight, forming an effective action in the presence of sources. We see that introducing the Wilson loop as observable actually amounts to a reweighting of the simulated system without static sources to a system with static sources present at exactly those spatial positions, at which the temporal stretches of the Wilson loop start and end. It is this possibility to introduce a posteriori the sources into the system, that makes conventional lattice QCD simulations so versatile. One set of numerical simulations can be reused to investigate multiple different physics scenarios, simply by a choice of observable.

The situation is quite different in the classical statistical approximation. There, we do not evaluate the path integral itself but instead evolve fields via deterministic classical equations of motion, while their initial values are drawn from a statistical ensemble. The first difference, we must note, is that computing an observable on the field configurations evolved in that way is not the same as evaluating the observable inside the full path integral. Hence, evaluating the Wilson loop on the classical equations of motion does not amount to the same reweighting we have seen taking place in the full quantum theory. I.e., evaluating the Wilson loop here leaves the gauge fields still oblivious of the presence of the static sources. 

As has been shown in detail in \cite{kasper_fermion_2014}, the classical equations of motion emerge naturally in the derivation of the classical statistical approximation from the full path integral. We may write the classical limit of the partition function for the gauge fields in the presence of fermions, denoted by $Q(x)$, as $Z_{\rm src}$ in the following form
\begin{align}
    Z_{\rm src}&=\int{\cal D}A(t=0) \int{ \cal D} \Pi(t=0)\rho(A,\Pi,t=0)\delta\Big(D_{\mu} F^{\mu\nu}[A]-j^\nu\Big)\,.\label{eq:partitionfunction}
\end{align}
What is left of the full path integral is an integration over the statistical distribution $\rho$ of the classical field $A$ and its conjugate momenta $\Pi$ at initial time. The dynamics of the fields at later times are determined by the classical equation of motion, housed in the delta function term, featuring the covariant derivative $D_\mu=\partial_\mu-igA_\mu$.

The coupling of fermionic degrees of freedom to the gauge fields in the classical statistical approximation introduces a current $j_\mu^a$, coupled to the classical gauge field degrees of freedom. This current is intimately related to the fermion propagator
\begin{align}
    j_\mu^a(x) =  \frac{g}{2} {\rm Tr}\big[ \langle [\bar Q(x), Q(x)] \rangle_{A} \gamma_\mu T^a\big] \overset{T/m_Q \ll 1}{=}\frac{g}{2} {\rm Tr}\big[ \langle \bar Q(x) Q(x) \rangle_{A} \gamma_\mu T^a\big] .
\end{align}

The gauge coupling $g$ appears together with the generators of the gauge group $T^a$ and the gamma matrices $\gamma_\mu$, required for coupling the fermions to the gauge field. For heavy fermions, the states, over which the expectation value is taken, do not contain any heavy quark particles. Thus, the commutator reduces to the forward correlation function where only creation operators act on the states from the left. Applying the Foldy-Tani-Wouthuysen transform to separate the contents of the four component Dirac spinor $Q=(\psi,\chi)$ into two two-component Pauli spinors, one finds that only the zeroth component of the current remains relevant. Its expression further simplifies to 
\begin{align}
    j_0^a(x) = \frac{g}{2} {\rm Tr}\big[\langle  \psi^\dagger(x) \psi(x) \rangle_{A}T^a - \langle  \chi^\dagger(x) \chi(x) \rangle_{A} T^a\big] + {\cal O}(m_Q^{-1}).
\end{align}
All other spatial components of this current are suppressed and therefore the information of the static sources only enters via the color density $j^a_0$. It makes its appearance in the classical theory as a source term on the RHS of the Gauss's law
\begin{align}
    D^a_iF_{0i}^{ab}=j_0^b(x),
\end{align}
written here concisely using the color components of the covariant derivative $D_\mu=\partial_\mu - igA_\mu$ and the field strength tensor $F_{\mu\nu}=[D_\mu,D_\nu]$. Being interested in the physics of color singlet states, we initialize the system with two static sources placed at ${\bf x}_1$ and ${\bf x}_2$ and select a color combination encoded in the $3\times 3$ matrices $M_1$ and $M_2$. We arrive at the proper Gauss's law for the gauge fields in the presence of two static sources in the classical statistical approximation
\begin{align}
    D^a_iF_{0i}^{ab} \overset{m\to\infty}{=} \frac{g}{2} {\rm Tr} \Big\{ T^b  \big( M_1 \delta^{(3)}({\bf x}-{\bf x}_1)+M_2\delta^{(3)}({\bf x}-{\bf x}_2) \big)\Big\}\label{eq:contgausslaw}\,.
\end{align}
The matrix' entries may be ordered as "red, green, blue". For a color-anticolor structure relevant to describe an overall color singlet state, we may then choose a red anti-red $r\bar r$-configuration, which we would represent by $M_1={\rm diag}[+1,0,0]$ and $M_2={\rm diag}[-1,0,0]$. This proper Gauss's law, as part of the equations of motion, implements the back-reaction of the sources onto the gauge fields. In turn, missing the source term amounts to neglecting this back-reaction all together.

As we will see in the following sections, this fact allows us to understand the absence of a real part in the potential observed in Ref. \cite{Laine:2007qy}. In that study, the propagating static color sources, described by the Wilson loop, did not lead to a back-reaction onto the gauge fields, which in turn did not allow them to build up a force among them. On the other hand, scattering of the medium gluons with the heavy quarks does not require such a back coupling and thus a finite imaginary part was found. 

In the present study, we will carry out simulations of the Wilson loop in classical statistical gauge theory in the presence of sources, i.e., based on the discretized counterpart of the proper Gauss's law of \cref{eq:contgausslaw}. It allows us to reveal the presence and values of the real part of the static potential. In addition, we investigate whether the back-reaction changes the behavior of the imaginary part of the potential found in Ref. \cite{Laine:2007qy}.

\section{Numerical Methods}
\label{sec:nummethods}

Let us briefly recollect the standard techniques of classical statistical lattice gauge theory for the gluon field degrees of freedom. 
Our starting point is the anisotropic Wilson plaquette action \cite{Wilson_1974,Klassen_1998} in the presence of a static charge density $j_0^a(x)$. We place it on a hypercubic lattice with spatial and temporal lattice spacing $a_s$ and $a_t$ respectively
\begin{align}
\nonumber S[U]=\frac{1}{g^2}\sum_{t}a_t \sum_{x} a_s^3\Big(&\sum_i \frac{2}{(a_sa_t)^2}{\rm ReTr}\big[ 1 - P_{0i}(x)\big] \\
- &\sum_{i,j} \frac{1}{a_s^4}{\rm ReTr}\big[ 1 - P_{ij}(x)\big] + A_0^a(x)j_0^a(x) \Big)
\end{align}
The plaquettes are defined as elementary Wilson loops $P_{\mu\nu}(x)=U_\mu(x)U_\nu(x+\hat\mu)U^\dagger_\mu(x+\hat\nu)U^\dagger_\nu(x)$ and are composed of the gauge links $U_\mu(x)={\rm exp}[ia_\mu g A^a_\mu(x)T^a]$, which contain the discretized gauge field $A^a_\mu(x)$ and the generators of SU(3), denoted by $T^a$. In the classical theory, the gauge coupling $g$ can be absorbed into a redefinition of the gauge fields and will therefore be set to unity in the following. The plaquettes are related to the gauge field strength tensor $P_{\mu\nu}(x)={\rm exp}\big[ ia_\mu a_\nu F_{\mu\nu}^a T^a]$ and in turn allow us to explicitly define the electric fields as $E_i(x)=E^a_i(x)T^a=F_{0i}^a(x)T^a$.

In order to simulate this system in a reliable fashion, we go over to a Hamiltonian formulation, where only space remains discretized and time becomes continuous. The strategy aims at formulating the equations of motion in canonical form, which in turn enables the deployment of symplectic time-stepping prescriptions, such as the ${\cal O}(\Delta t ^2)$ accurate leap-frog algorithm for separable Hamiltonians. The canonical form of the equations of motion are obtained by choosing a suitable gauge. In temporal gauge with $U_0(x)=1$, the dynamical degrees of freedom are the gauge group valued spatial links $U_i(x)$ and their generator valued conjugate momenta $E^a_i(x)$, residing on individual time slices. The Hamiltonian according to the plaquette action reads
\begin{align}
    \nonumber H &= 
    \Big[ L_E - L_M \Big]\\
    \nonumber &= \sum_x \Big\{   \sum_{i,a} \frac{a_s^3}{2}\big( E^a_i({\bf x},t) \big)^2 + \sum_{i,j} \frac{1}{a_s}{\rm ReTr}\big[ 1 - P_{ij}({\bf x},t)\big] \Big\}\\
    &= \frac{1}{a_s}\sum_x \Big\{   \sum_{i,a} \frac{1}{2}\big( a_s^2 E^a_i({\bf x},t) \big)^2 + \sum_{i,j} {\rm ReTr}\big[ 1 - P_{ij}({\bf x},t)\big] \Big\} = \frac{1}{a_s}\bar{H},
\end{align}
where in the last line we have introduced the dimensionless $\bar{H}$, explicitly recovering that the units of $H$ are that of the inverse spatial lattice spacing.
The dynamics of the gauge fields is described via the spatially discretized e.o.m.
\begin{align}
    &\dot{U}_i({\bf x},t)=ia_sE^a_i({\bf x},t)T^a U_i({\bf x},t),\\
    &\dot{E}_i^a({\bf x},t)=\frac{2}{a_s^3}{\rm ImTr}\Big[\sum_{j\neq i} T^a U_j({\bf x},t). (S^\sqsubset_j+S^\sqsupset_j)\Big] \label{eq:hamevol}
\end{align}
Here, $S^\sqsubset_j$ and $S^\sqsupset_j$ denote the backward and forward staple, which are the leftovers of the full backward and forward plaquettes where the link $U_j$ has been removed, i.e., they yield the backward and forward plaquette once multiplied with the link $U_j$ from the right or left respectively. In addition to these equations of motion, the functional derivative of the lattice action with respect to the $A^0$ component of the gauge field leads to the Gauss's law in the presence of static sources
\begin{align}
    G({\bf x},t)=\frac{1}{a_sa_t}\sum_i \Big[ E_i({\bf x},t)- U_i^\dagger({\bf x},t) E_i({\bf x},t)U_i({\bf x},t)\Big]- \frac{1}{a_t}j^a_0({\bf x})T^a = 0 \label{eq:gausslaw}.
\end{align}
If consistent initial conditions are chosen such that they fulfill \cref{eq:gausslaw}, then the evolution equations \cref{eq:hamevol} will preserve this property. When implemented as discrete time stepping, the dynamics will lead to small deviations from Gauss's law, which we check to remain insignificant at all times.

In order to generate the collection of stochastic initial conditions according to the classical density matrix at temperature $T=1/\beta$, we resort to the strategy originally devised in \cite{Ambjorn:1990pu,Ambjorn:1997jz,Grigoriev:1988bd} and its implementation as outlined in e.g. \cite{Akamatsu:2015kau}. 

The thermal probability distribution of the degrees of freedom follows the Boltzmann weight $P[E,U] \propto {\rm exp}\Big[ -\beta H[E,U] ]$. As the Hamiltonian is separable, the electric field contribution is independent of the links and it appears at first sight that the $E_i^a$s are simply Gaussian distributed. Gauss's law however distinguishes among the electric fields those that are physical and those that are not. Hence we will have to not only stochastically draw values of $E_i^a$ according to
 \begin{align}
  E^a_i=\eta^a_i,\quad \langle \eta^a_i \rangle =0,\quad \langle \eta^a_i\eta^b_k \rangle= \sigma^2 \delta_{ab}\delta_{ik},\quad \sigma^2=\frac{1}{a_s^3 \beta},\,
 \end{align}
but in addition project to the physical subspace of electric fields by subsequently minimizing, via gradient descent, the functional $\sum_{x,a}{\rm Tr}[T^a G({\bf x},t) ]^2$. Note that it is only here in the projection of the initial conditions where the effect of the static sources enters the gauge field dynamics, as it is here where the Gauss's law is enforced. 

With a first set of projected electric fields at hand, we may proceed to compute the corresponding spatial links. To this end, we evolve some arbitrary initial set of links according to \cref{eq:hamevol} based on the quasi-thermal electric fields found above. This allows them to mutually equilibrate with a temperature, which however is not yet the desired one. To reach the thermal steady state between $E_i^a$ and $U_i$, defined by $\beta$, the electric fields need to be redrawn, projected and evolved with the links several times. The outcome of this procedure in turn is used as one of the realizations of the thermal initial conditions for the actual real-time evolution of the gauge fields. Since the steady state is characterized by a constant energy, we have made sure that the mutual equilibration between links and electric fields is realized well enough, so that any remaining changes in the energy are below percent level.

In temporal gauge the computation of the Wilson loop simplifies considerably, as the temporal links are unit matrices. Due to time translational invariance in thermal equilibrium, we compute the real-time Wilson loop spanning from the initial time slice to the time slice, which the simulation has currently reached. To this end we keep a copy of the initial conditions in memory and form the following products
\begin{align}
    W^{j}_{\bf x}(r,t) = {\rm Tr}\Big[ \prod_{l=0}^{r/a_s} U_j({\bf x}+a_s l \hat j,0) \Big( \prod_{l=0}^{r/a_s} U_j({\bf x}+a_s l \hat j,t) \Big)^\dagger \Big]\,.
\end{align}
In case that the sources are set to zero, the Wilson loop expectation value does not depend on the axis direction and position of it starting point so that $\langle W_\square ^{\rm no\, src}(r,t)\rangle= \langle \sum_{\bf x} \sum_j W^{j}_{\bf x}(r,t) \rangle/\sum_{\bf x} \sum_j 1$. On the other hand in the presence of sources, we choose to place one of them at the origin and the second one along the x-axis at distance $r_{\rm src}$, such that the Wilson loop at only a single position and with a single spatial extent is computed over time $ \langle W_\square^{\rm src}(r_{\rm src},t)\rangle = \langle W^{x}_{\bf 0}(r_{\rm src},t)\rangle$. The lack of geometric averaging in this case leads to significantly larger statistical uncertainties in the observable, which in turn need to be compensated for by collecting more statistics for the ensemble average over initial conditions.

As discussed in \cref{sec:introduction}, the interaction potential between the static charges may be computed using the spectral function of the Wilson loop. While in standard Euclidean lattice simulations this requires to solve an ill-posed inverse problem, in a real-time simulation we only have to carry out a Fourier transform. To minimize the computational cost, we exploit that the Wilson loop at positive and negative times is related by complex conjugation $W_\square(t,r)=W^*_\square(-t,r)$ in order to carry out a discrete Fourier transform on its values in an interval $[-t_{\rm max},t_{\rm max}]$. Since the discrete Fourier transform is defined for periodic signals, it is paramount that the simulation has progressed far enough so that the values of the Wilson loop at the latest time have decayed close enough towards zero. This is realized in most cases except for the smallest spatial separation distances which may lead to ringing artefacts in the resulting spectral functions. The further the amplitude of the signal has decayed over time the weaker the ringing becomes. As is well known, such ringing can be taken care of by introducing appropriate windowing functions. We deploy the well established Hann window function \cite{Hann_window} to the real-time correlation functions at the lowest and next to lowest spatial separations in order to avoid such ringing artifacts. We have checked that this procedure does not introduce a bias onto the estimation of the potential beyond the statistical uncertainties.

With the spectral function of the Wilson loop at hand we are in a position to investigate the binding properties of static color charges in lattice Yang-Mills theory in classical thermal equilibrium. 

\section{Numerical Results}
\label{sec:numres}

The main results of our study are computed based on lattices with $N=32$ grid points in each spatial direction (The source code for our simulations is freely available under an open access license, hosted at the Zenodo repository \cite{alexander_rothkopf_2020_4332406}). We set the spatial lattice spacing to unity $a_s=1$ and work with a time stepping $\Delta t=0.1$. Each realization of the microscopic classical dyanmics, started from an independent stochastically drawn set of $N_{\rm runs}$ initial conditions, is evolved up to times $N_t \Delta t=1200$, both in the absence and in the presence of sources. To generate the thermal initial conditions we draw and project the electric fields $N_{\rm init}=20$ times and each time evolve them with the gauge fields for $N_{\rm therm}=150$ steps using the above $\Delta t$. Supplementary plots showcasing the initialization as well as the conservation of the Gauss's law and energy can be found in \cref{sec:AppSupplementNum}.

\begin{table}[h]
\centering
\begin{tabular}{lllll}
$\beta$ & $a_s$ & src &$N_{\rm runs}$ &  \\ 
\hline\hline
4 & 1 & NO  &500 &    \\
6 & 1 & NO  &200 &   \\
8 & 1 & NO  &80 &   \\
10& 1 & NO  &80 &   \\
12& 1 & NO  &80 &   \\
16& 1 & NO  &80 &   \\
20& 1 & NO  &80 &   \\
20& 1 & YES  &1200 &   \\
24& 1 & NO  &80 &   \\
28& 1 & NO  &80 &   \\
32& 1 & NO  &80 &   \\
\end{tabular}
\caption{Parameters used in the simulations underlying our extraction of the proper static interaction potential in classical lattice gauge theory.}
\end{table}

\begin{figure}
\includegraphics[scale=0.5]{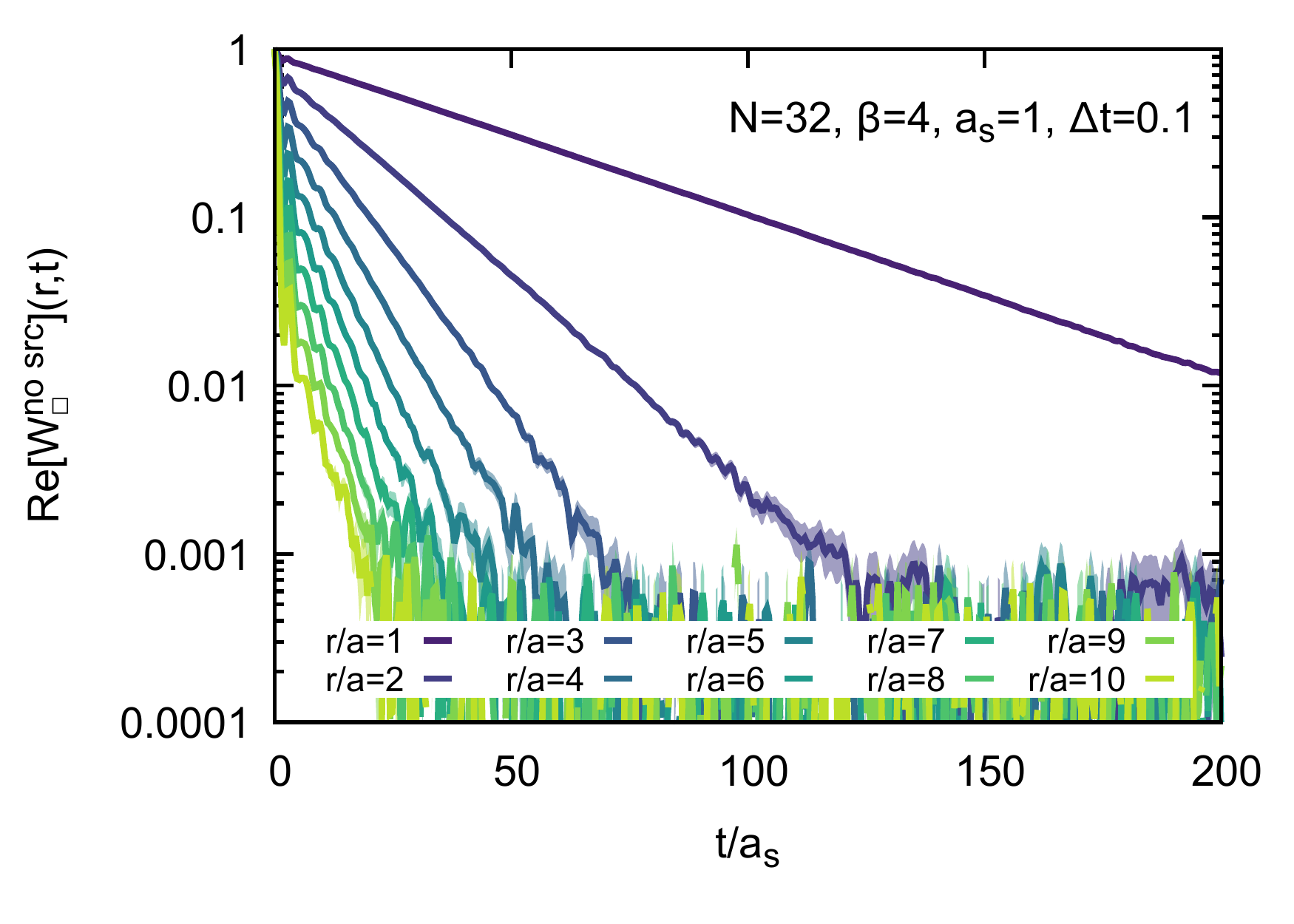}
\includegraphics[scale=0.5]{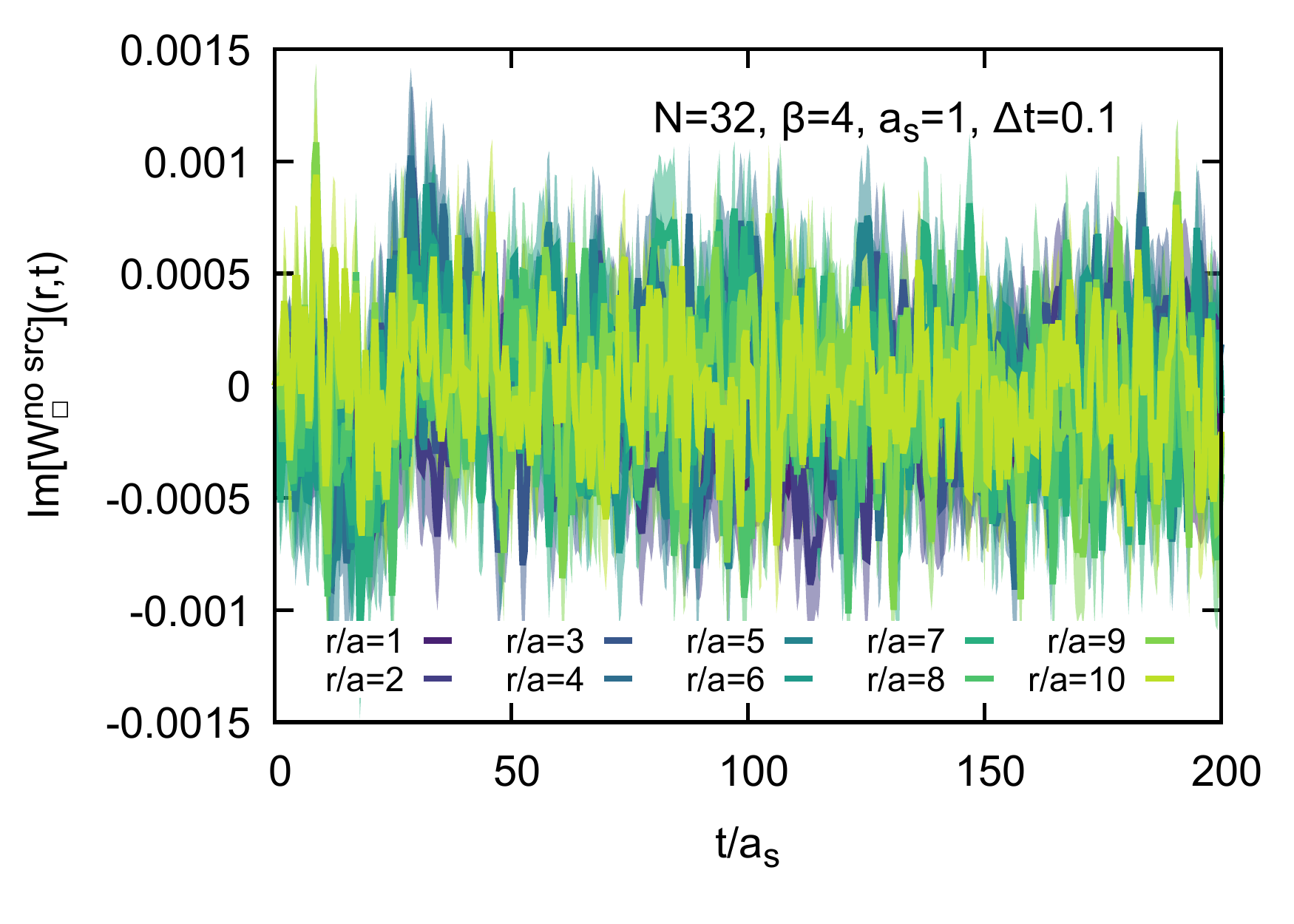}
\caption{Real- (top) and imaginary (bottom) part of the real-time Wilson loop evaluated at different spatial on-axis separation distances between $r=1\ldots10a_s$ on $N_x=32$ lattices with $\beta=4$ in the absence of a back-reaction between static sources and the surrounding medium. The real part of the Wilson loop is positive and exhibits clear exponential damping, while the imaginary part is compatible with zero.}\label{fig:NoSrcWLoop}
\end{figure}

\subsection{Simulations without explicit source term}

As a first step, we reproduce and refine the results of \cite{Laine:2007qy}, in which the thermal real-time Wilson loop was studied in the absence of an explicit source term in Gauss's law. In order to resolve the changes due to variation in the system temperature, we choose ten values of $\beta=1/Ta_s$ between $4$ and $32$. Using slightly different conventions than in \cite{Laine:2007qy}, our choice of $\beta=4$ corresponds to $\beta_{\rm L}=16$ in the work of Laine and Tassler. As a representative example, we plot in \cref{fig:NoSrcWLoop} the Wilson loop's expectation value at $\beta=4$ along real-time $t$ evaluated at several different spatial distances $r/a_s=1\ldots10$ (darker to lighter lines). $\beta=4$ corresponds to the highest temperature considered in this study and the falloff of the Wilson loop is thus the most rapid, requiring $500$ independent initial conditions to reach the level of statistical uncertainty, indicated by the errorbands.

The top panel of \cref{fig:NoSrcWLoop} displays the real-part of $\langle W^{\rm no\,src}_\square(r,t)\rangle$, while the bottom panel shows its imaginary part. As expected from Ref.\cite{Laine:2007qy}, the Wilson loop in the absence of static sources is purely real within statistical uncertainty. A first inspection of this logarithmic plot by eye indicates that at late times it decreases approximately as a single exponential. It is this exponent that has previously been identified with the imaginary part of the potential between static quark--anti-quark sources. 

\begin{figure}
\includegraphics[scale=0.5]{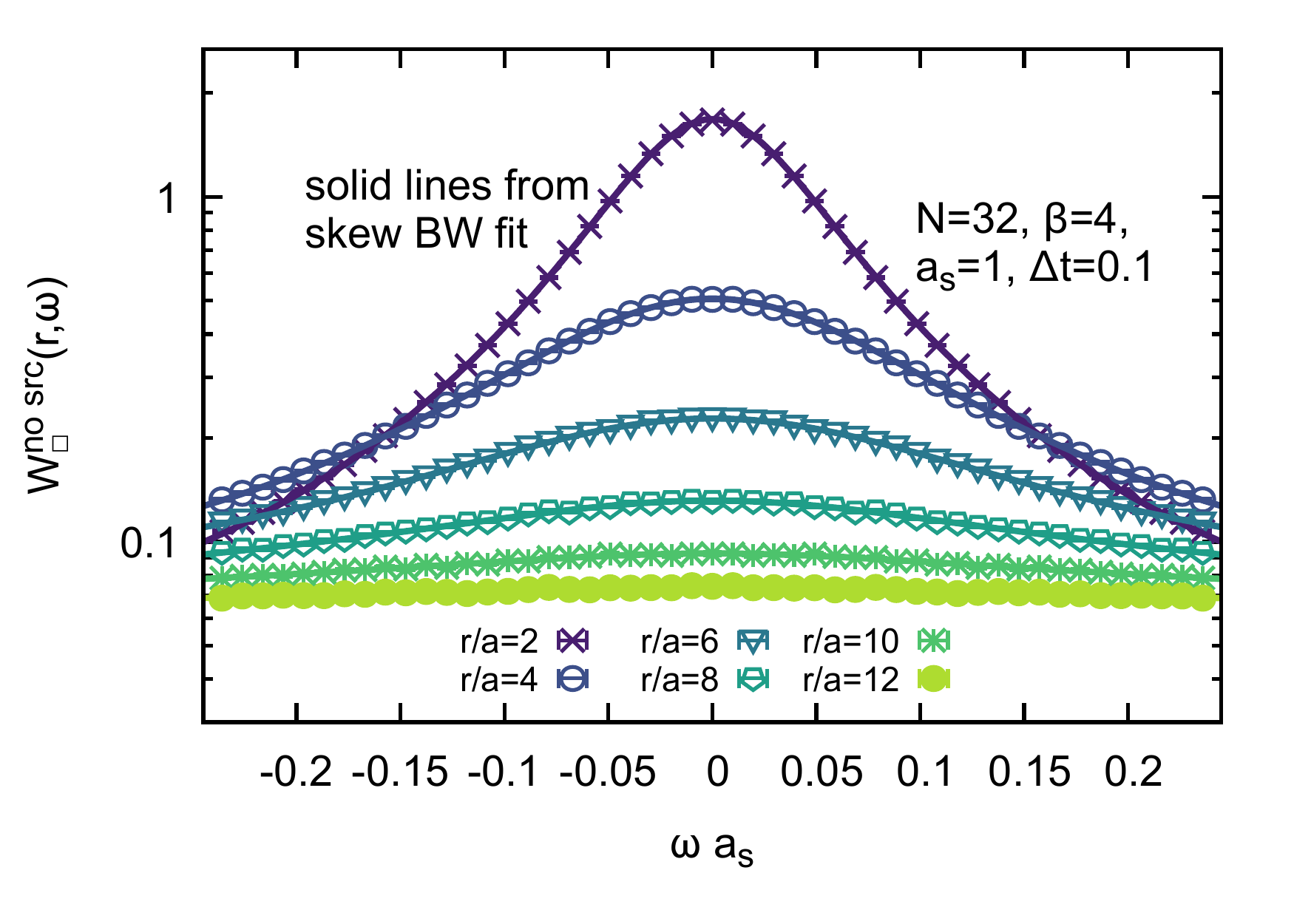}
\caption{Discrete Fourier transform (colored points) of the real-time Wilson loop evaluated at different spatial on-axis separation distances between $r=1\ldots12a_s$ on $N_x=32$ lattices with $\beta=4$ in the absence of a back-reaction between static sources and the surrounding medium. Solid lines indicate the best fit result of the dominant peak structure at zero frequencies.} \label{fig:NoSrcSpectralFunc}
\end{figure}

In order to extract the potential from the Wilson loop, we proceed to scrutinize its spectral function, computed via discrete Fourier transform, plotted as individual datapoints in \cref{fig:NoSrcSpectralFunc} for six different spatial separation distances $r/a_s=2,4,6,8,10$ and $12$. We plot as range of frequencies $\omega a_s \in [-0.25,0.25]$, where the relevant structure encoding the potential is present. The full frequency range extends to the much larger values given by $\pi/\Delta t$ (not shown). In agreement with \cref{fig:NoSrcWLoop}, the spectral function is purely real and and exhibits a single dominant peak located at vanishing frequency, whose width increases with spatial separation distance of the underlying coordinate space Wilson loop.

\begin{figure}
\includegraphics[scale=0.5]{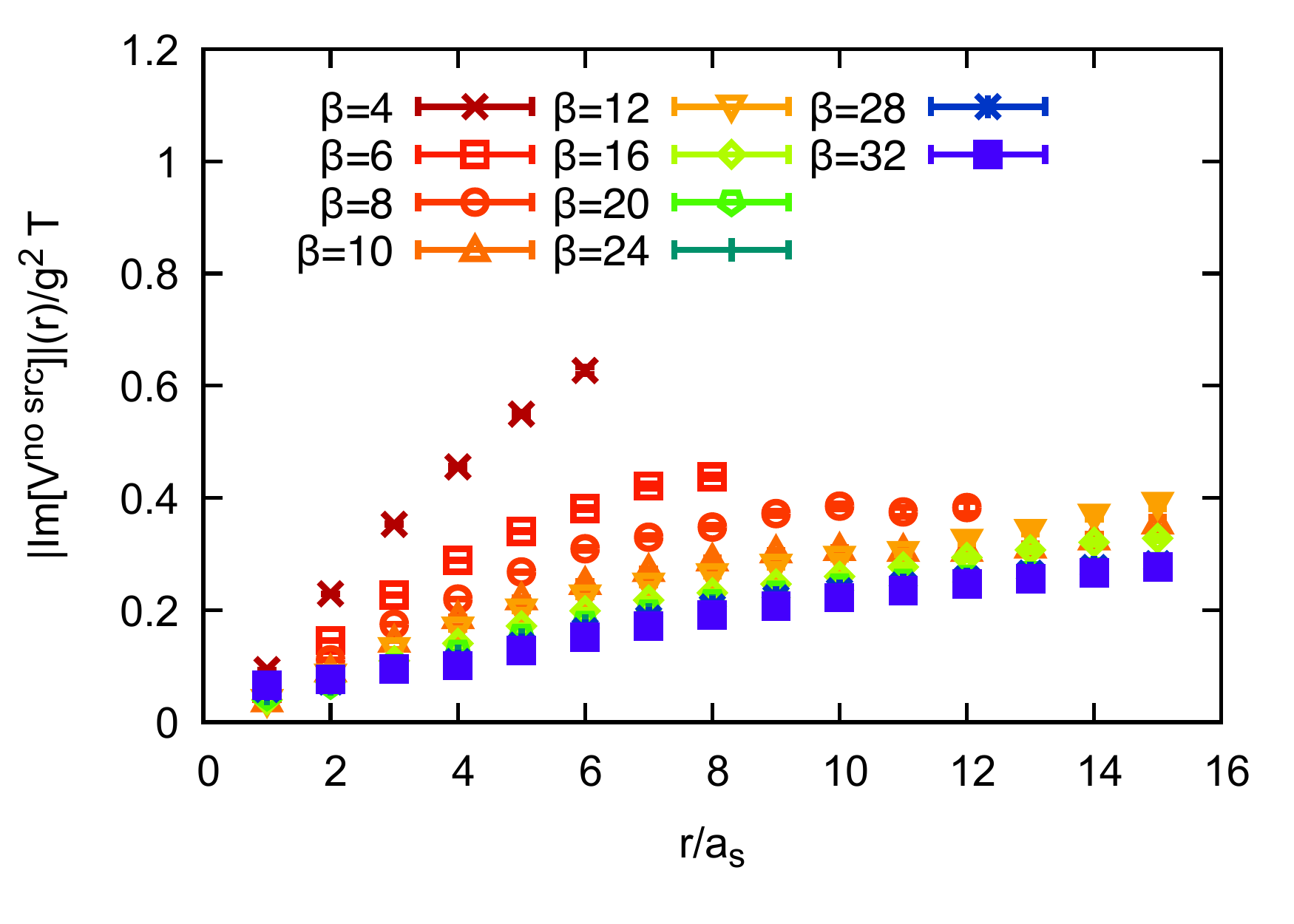}
\includegraphics[scale=0.5]{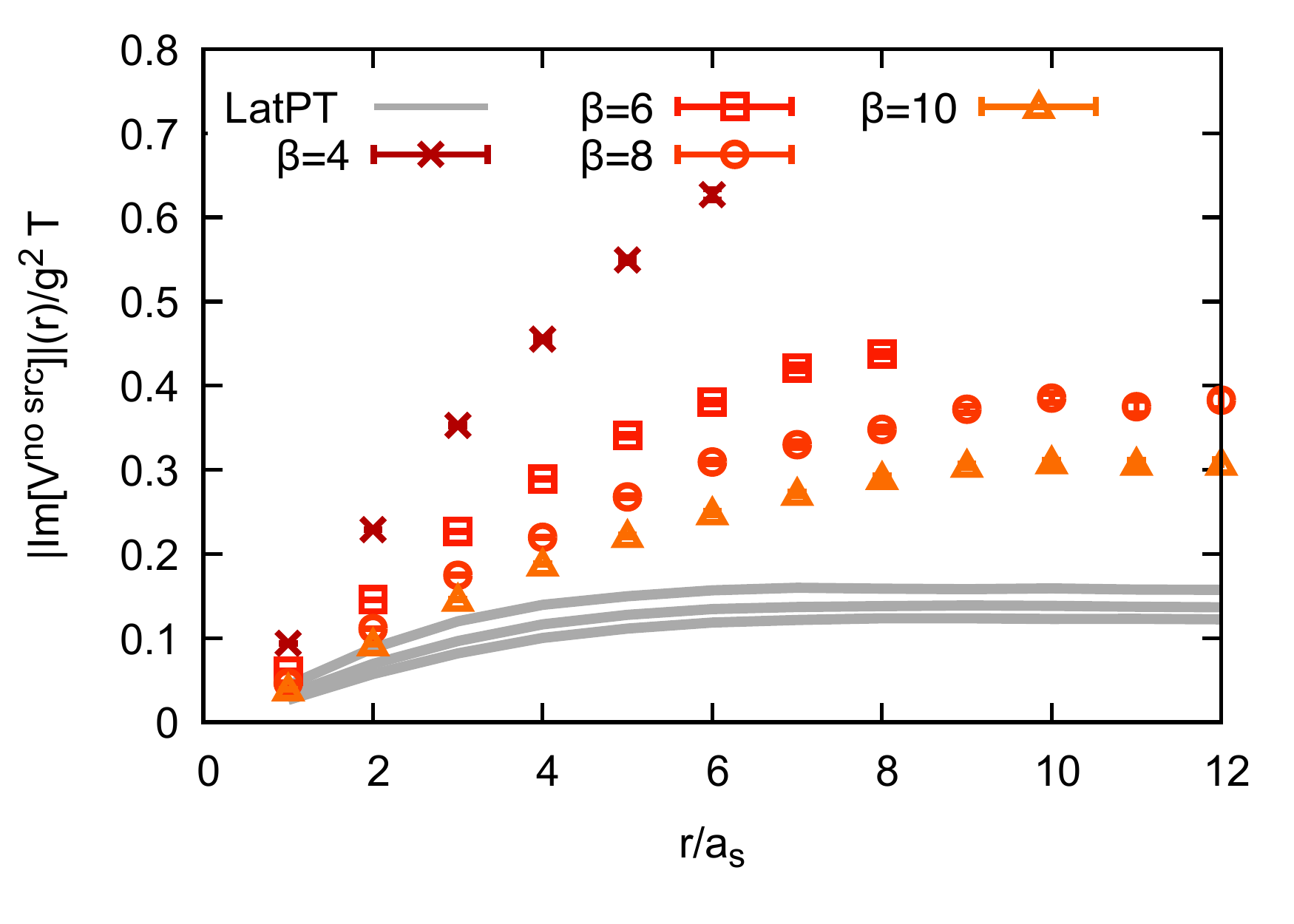}
\caption{Real- (top) and imaginary part of the static quark potential evaluated at different temperatures $\beta=4\ldots 32$ on $N_x=32$ lattices in the absence of a back-reaction between static sources and the surrounding medium. (bottom) comparison of the results at the four highest temperatures to the corresponding values from lattice perturbation theory.} \label{fig:NoSrcPot}
\end{figure}

The solid lines correspond to the best fit according to \cref{eq:sBWfit}, using 15 of the discrete points of the spectral function above and below the peak maximum. The skewed Breit-Wigner shape predicted on general grounds in \cite{Burnier:2012az} matches the data very well with $\chi^2/{\rm d.o.f.} \approx 1.5\ldots2$. Changes in the fitting range do not significantly change its outcome. In order to assign uncertainties to the extracted values of the potential we have deployed a ten bin Jackknife, whose variance underlies the relatively small errorbars shown in \cref{fig:NoSrcPot}.

In order to crosscheck the results obtained from the spectral function analysis, we have also carried out exponential fits to the late time behavior of the coordinate space Wilson loop. We see that at small separation distances $r/a<6$ where the single exponential decay of the Wilson loop is well resolved, both methods agree. At larger spatial separation distances, it becomes more and more difficult to select by eye an appropriate fitting regime, leading to exponential fit values for ${\rm ImV}$ that are slightly larger than those from the spectral function analysis, an effect we attribute to excited state contamination. Considering the agreement with the exponential fit at small distances and the fact that the DFT based extraction shows very small dependence on the fitting range, we deem it reliable enough in the context of the currently available input data quality.

Tabulating the values of the peak position and width for different separation distances provides us with an estimate of the static quark interaction potential in the absence of a back-reaction of those sources onto the gauge fields. Carrying out the same procedure at different values of $\beta$, we elucidate the temperature dependence of the potential which is plotted in the top panel of \cref{fig:NoSrcPot}. What we obtain from the width are positive numbers. These correspond to the magnitude of the imaginary part. Its values are surely negative, as they lead to a dampening of the amplitude of the Wilson loop over time. With the imaginary part in the full quantum theory possessing a trivial dependence on temperature, we plots here its values divided by the system temperature.

We find an imaginary part, which increases monotonously with temperature. Based on the state-of-the-art extraction procedure using the Wilson loop spectral functions, the values of ${\rm Im}[V]$ obtained here are robust at the current level of statistical uncertainty up to  $r/a_s=16$ for $\beta \geq 16$, while the loss of signal to noise ratio limits the extraction to smaller distances for lower values of $\beta$. 

Qualitatively similar to the predictions of HTL perturbation theory in the fully quantum case, we find two distinct regions. One, at small distances (here $r/a_s<8$), which features a relatively steep increase and a convex shape. The other at subsequently larger distances, with a much weaker slope and a concave behavior that appears to lead to a flattening of the values of ${\rm Im}V$ at larger distances.

We may compare the simulation results also quantitatively to the predictions from the classical limit of lattice perturbation theory. Leveraging equation (3.23) from Ref. \cite{Laine:2007qy}, we compute the values of $|{\rm Im}V|/g^2T$ for the $\beta$ values in our study (see the supplementary material for an explicit implementation in Mathematica). The outcome is plotted in the top panel of \cref{fig:NoSrcPot} as gray solid lines. Laine and Tassler evaluated the corresponding $\beta=4$ imaginary part for the first four spatial distances. We see that while the qualitative behavior appears similar, quantitative differences exist. In general the perturbative values are smaller than those obtained from the lattice. In addition the perturbative ${\rm ImV}$ reaches its asymptotic value at earlier distances compared to what we observe in the simulation. It is not surprising that also the large distance asymptotic value is smaller in the weakly coupled case, as $\rm{Im}V(r\to\infty)$ is related to single heavy quark energy loss in the medium. A strongly coupled environment intuitively induces energy loss more efficiently than a weakly coupled one.

\subsection{Simulation with explicit source terms}
\label{sec:evolsrc}
In this section we advance toward the main result of our study, the determination of the proper interaction potential between static color sources in classical lattice gauge theory at finite temperature. As discussed in \cref{sec:statsource} the missing ingredient in studies so far was the inclusion of the back-reaction of the static sources on the gauge fields. Now with source terms explicitly present, we have to carry out separate simulations for each of the different spatial separation distances we wish to observe the static color sources at. In addition, it is not possible to average the Wilson loop over axis-directions and spatial positions, which requires a factor of up to 10 times higher statistics in order to reach satisfactory results. We therefore restrict the study to one single temperature given by $\beta=20$. Our choice is motivated by the fact that at this $\beta$, in simulations without explicit source terms present, thermal effects are already visible, while at the same time a sufficient signal-to-noise ratio can be obtained. 

As discussed in \cref{sec:nummethods}, the novel element in this study is the treatment of the thermal initial conditions in order to take into account the change in Gauss's law due to the sources. We have checked explicitly (see \cref{sec:AppSupplementNum}) that this proper Gauss's law is preserved by the dynamical evolution.

\begin{figure}
\includegraphics[scale=0.5]{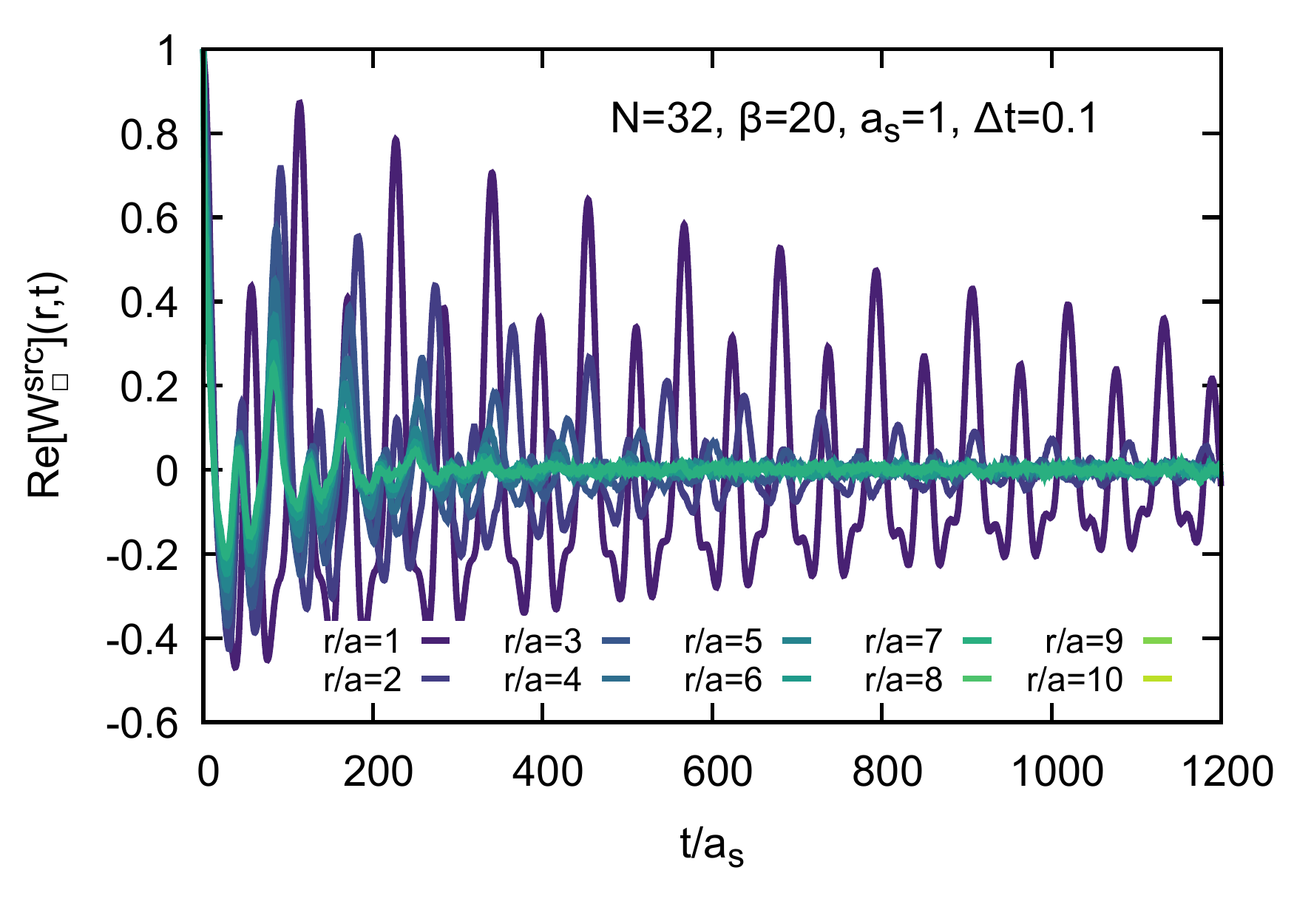}
\includegraphics[scale=0.5]{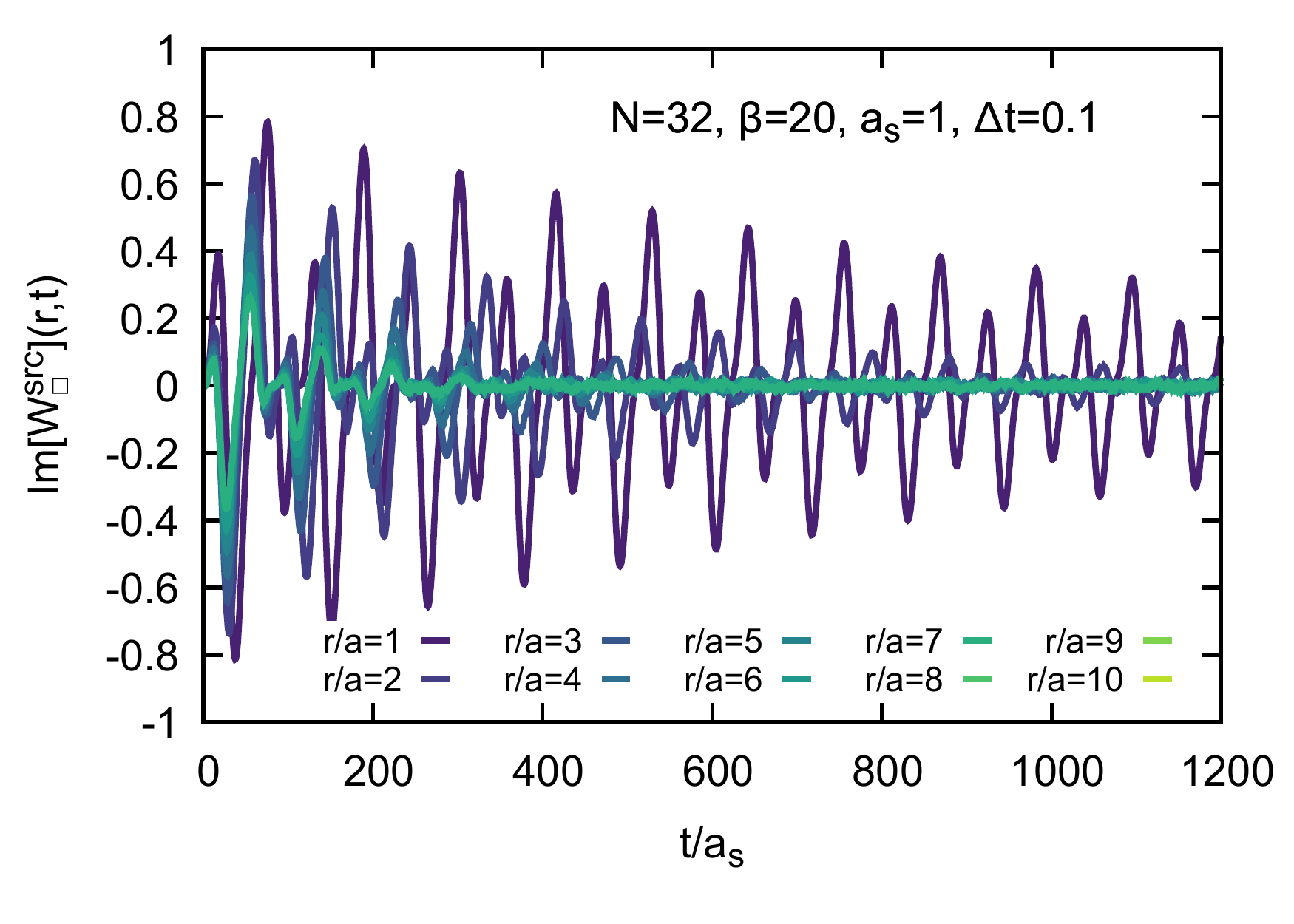}
\caption{Real- (top) and imaginary (bottom) part of the real-time Wilson loop evaluated with sources placed at different spatial on-axis separation distances between $r=1\ldots10a_s$ on $N_x=32$ lattices with $\beta=20$. In the presence of back-reaction, both quantities show finite values and exhibit a combination of oscillatory and damping behavior.}\label{fig:SrcWLoop}
\end{figure}

In \cref{fig:SrcWLoop}, we show the resulting real- (top) and imaginary part (bottom) of the Wilson loop at different spatial separation distances $r/a_s=1\ldots 10$ (darker to lighter solid lines) at inverse temperature $\beta=20$. The expectation values are computed from $N_{\rm runs}=1200$ simulations, carried out with independently drawn initial conditions. The simple modification in the initial conditions has fundamentally changed the behavior we observe. 

Not only does the imaginary part of the Wilson loop now show finite values, ${\rm Re}[V]$ and ${\rm Im}[V]$ both exhibit oscillatory behavior with a monotonously decreasing amplitude also. Note that the former starts from unity while the latter starts from zero at the origin, consistent with the transformation properties of the Wilson loop under time reversal. A first inspection by eye hints at the presence of at least two sinusoidal contributions to the dynamics. The exponential suppression of the amplitude of the Wilson loop with increasing spatial separation distances is also present here, which leads to a quickly deteriorating signal to noise ratio at increasing $r/a_s$. 

While the decrease in amplitude hints at the presence of a finite imaginary part of the potential, the oscillatory component bodes well to identify a real-part of the interaction potential as well. 

\begin{figure}
\includegraphics[scale=0.5]{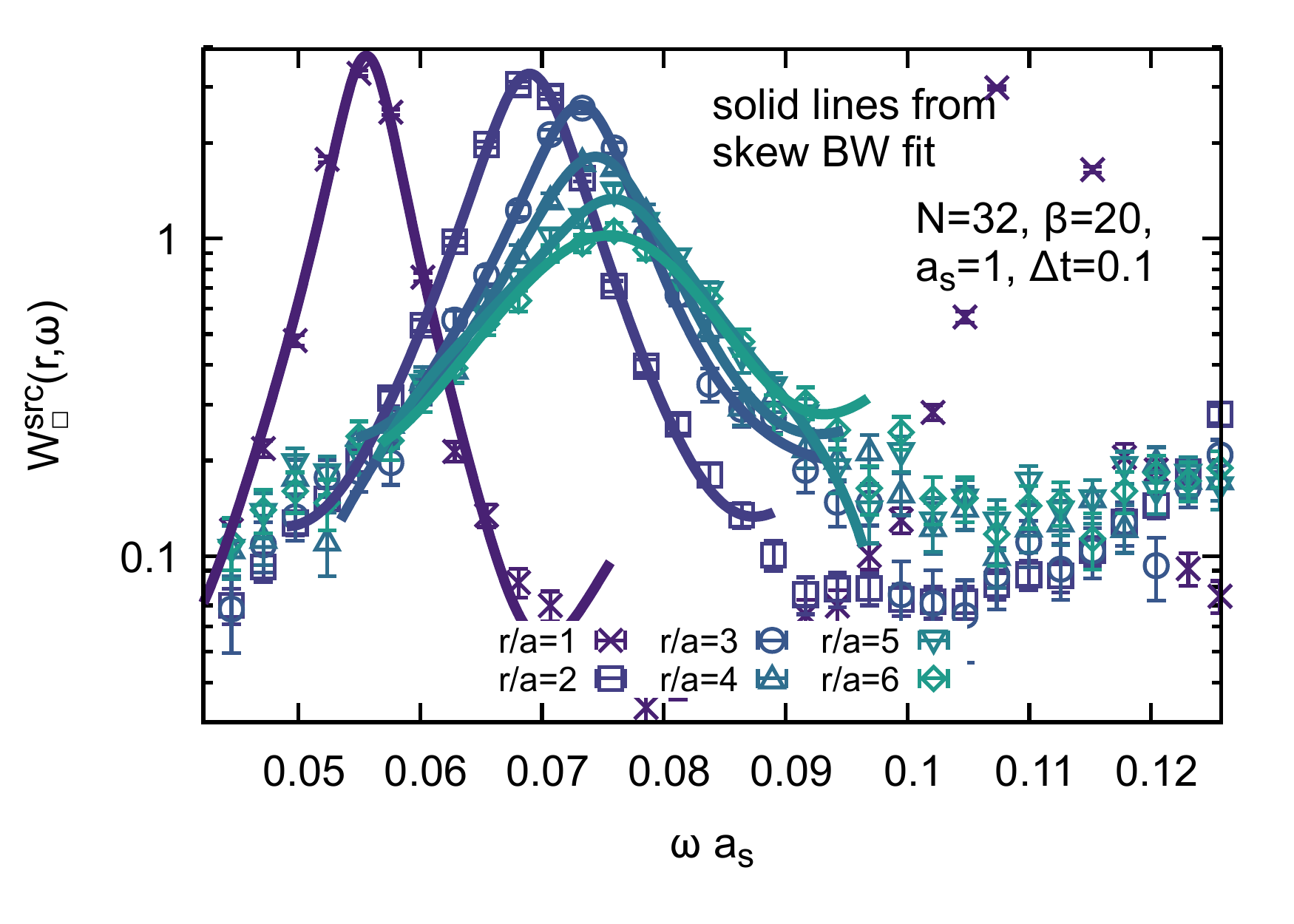}
\caption{Discrete Fourier transform (colored points) of the real-time Wilson loop evaluated in the presence of sources placed at different spatial on-axis separation distances between $r=2\ldots6a_s$ on $N_x=32$ lattices with $\beta=20$. For clarity only the positive frequencies of the mirror symmetric function are shown. Solid lines indicate the best fit result of the lowest lying peak structure around its maximum, relevant for the extraction of the static quark potential.} \label{fig:SrcSpectralFunc}
\end{figure}

Let us proceed to a quantitative extraction of the potential for which we first compute the spectral function of the Wilson loop via DFT. In \cref{fig:SrcSpectralFunc} we plot the corresponding values (colored symbols) for positive frequencies in the limited domain between $\omega a_s\in[0.045,0.12]$ where we locate one of the two dominant features of the spectrum, which is directly related to the potential. The results for six spatial separation distances are plotted between $r/a_s=1$ and $6$ (darker to lighter color).

Due to the significantly lower signal to noise ratio, the data points here show much more variation than in the case without sources. Since the amplitude of the $r/a_s=1,2$ Wilson loop has not yet decayed significantly at the final time $t a_s = 1200$ of our simulation, the naive Fourier transform would exhibit ringing around the lowest lying peak. We thus have applied the Hann windowing function in time to the coordinate space Wilson loop before the DFT. We find that also in the case with sources present, the lowest lying spectral structure, around its maximum, can be captured well with the skewed Breit-Wigner form of \cref{eq:sBWfit}, as shown by the best fit results as solid lines. Both the peak position and the peak width show a clear dependence on the separation distance. 

\begin{figure}
\includegraphics[scale=0.5]{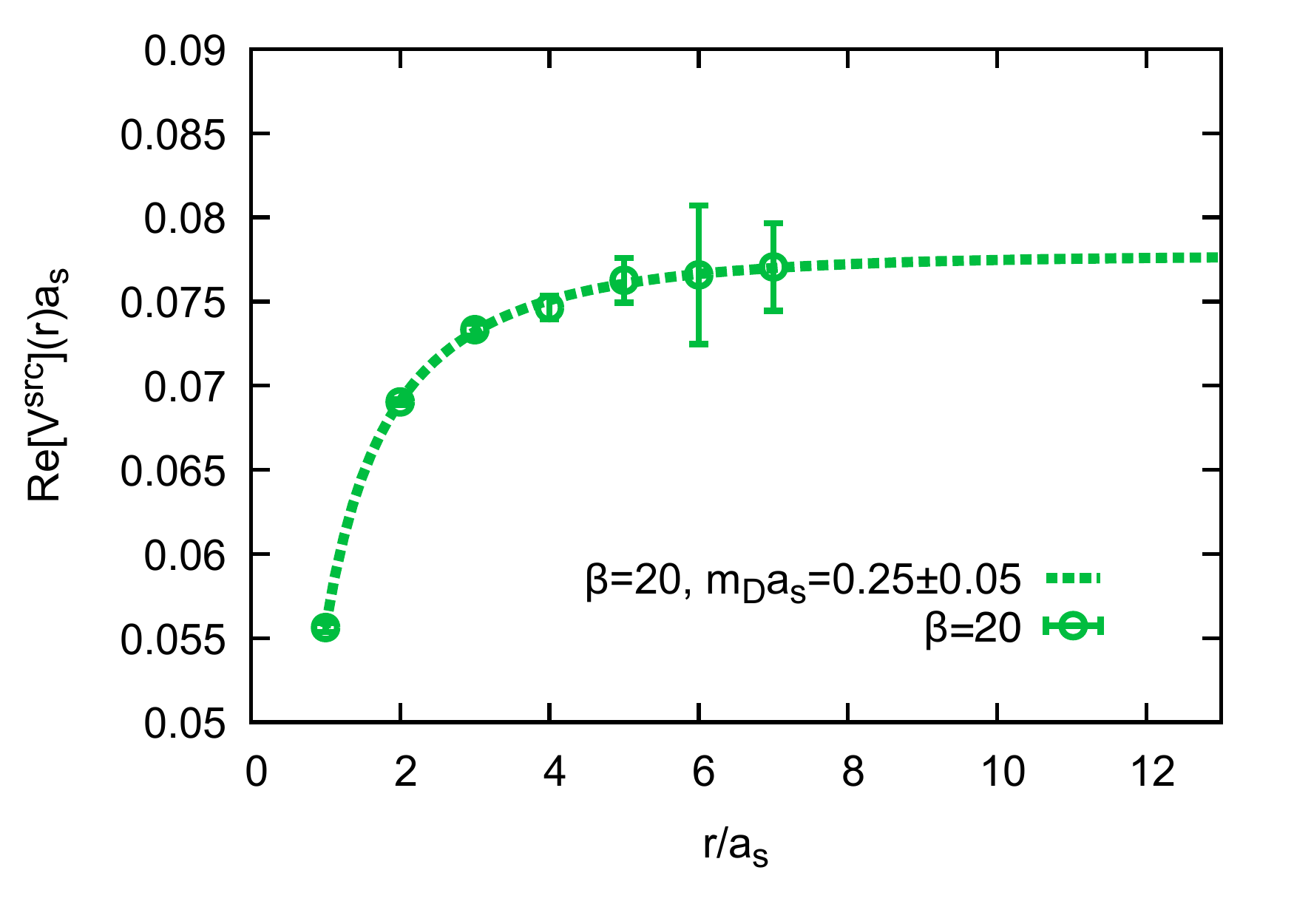}
\includegraphics[scale=0.5]{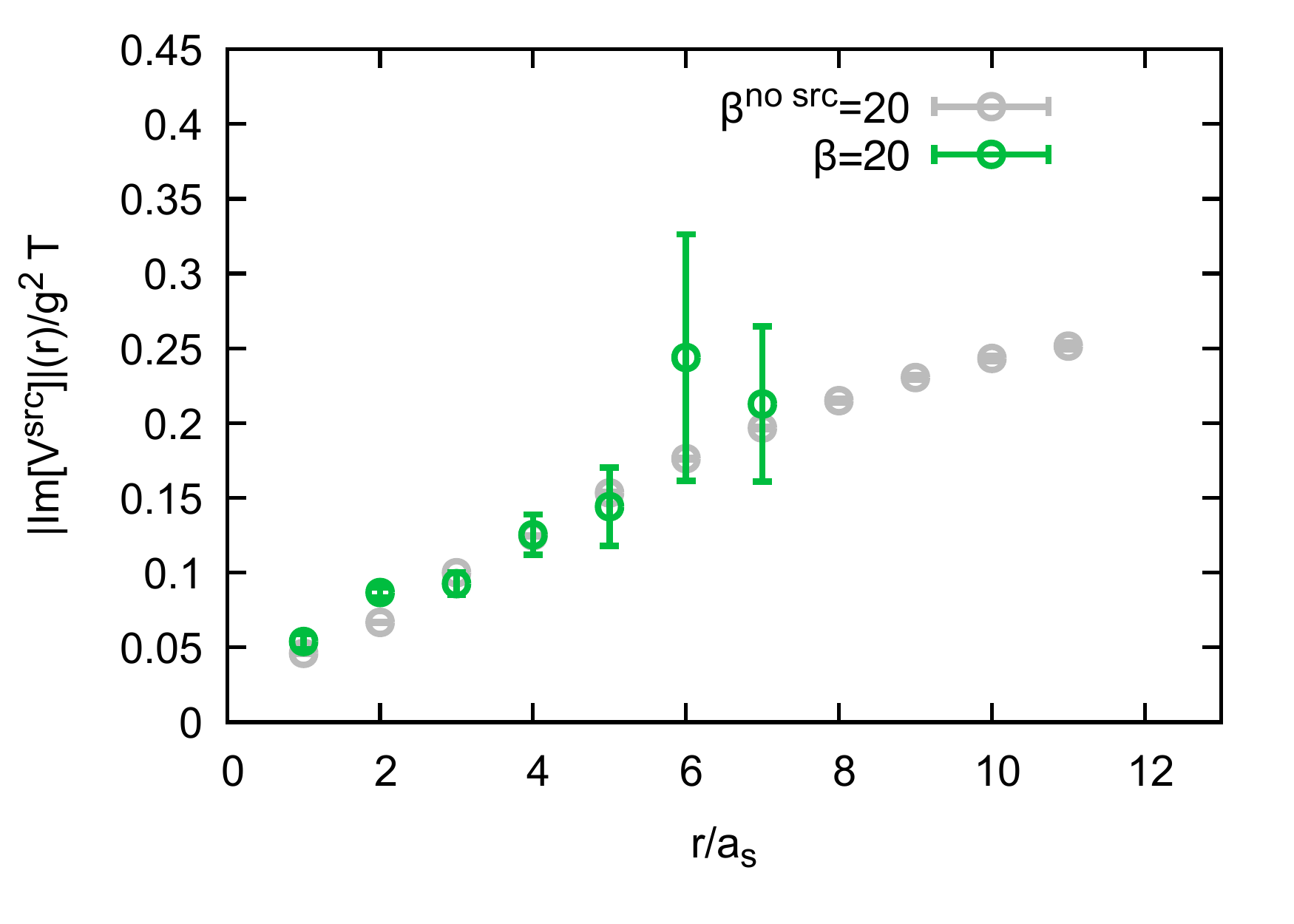}
\caption{Real- (top) and imaginary (bottom) part of the static quark potential  evaluated in the presence of sources at $\beta=20$ on $N_x=32$ lattices. As expected from the behavior of the real-time Wilson loop we now find both a finite real- and imaginary part. The former exhibits clear signs of screening as indicated by the Debye fit (dashed gray line). The latter shows the expected monotonous increase with spatial distance. Results in the absence of sources are shown as gray symbols and reveal a very good agreement of ${\rm Im}[V]$ between the simulations with and without back reaction.}\label{fig:SrcPotential}
\end{figure}

By determining the peak positions and widths at $\beta=20$, we arrive at the central result of this study, the values of the real- and imaginary part of the proper static interquark potential, shown in \Cref{fig:SrcPotential}. As expected by the presence of a finite imaginary part in the coordinate space Wilson loop $\langle W_\square^{\rm src}\rangle(r,t)$ and its damped oscillatory behavior, we do indeed find finite values for the real-part, as shown in the top panel of \cref{fig:SrcPotential}. The errors are dominated by the fit uncertainty of the seven-parameter $\chi^2$ fit according to \cref{eq:sBWfit}, which has been carried out for ten Jackknife bins.

Since classical lattice gauge theory does not capture the physics of confinement, one may expect to find a Coulombic behavior at very small distances corresponding to the tree-level short range interaction in Yang-Mills theory. Since we are observing the system at finite temperature, a hot medium of color charges fills the space between the static sources and in analogy with the well-known Abelian theory, we thus expect that the interactions are screened. And indeed using the simple ansatz $-\frac{A}{r}{\rm exp}[-m_Dr]+{\rm const.}$ it is possible to reproduce the dependence of ${\rm Re}[V]$ on the spatial separation distance $r$ very well (dashed green line). Such a Debye screened behavior is qualitatively very similar to what has been observed also in the fully quantum theory deep in the deconfined phase. 

The value of $m_D$ we observe is in good agreement with that predicted by lattice perturbation theory according to \cref{eq:DebyeMassLPT}: 
\begin{align}
& (m_Da_s)(\beta=20)=0.25(5) & (m_D^{\rm lPT}a_s)(\beta=20)=0.275
\end{align}

In the bottom panel of \cref{fig:SrcPotential}, we plot the absolute value of the imaginary-part of the proper potential as colored symbols, again divided by the temperature. A very similar picture emerges as in the simulations without explicit sources. The imaginary part increases monotonously with separation distance. Plotting the results from simulations without explicit inclusion of sources at the same temperature as light gray symbols, we actually find very good agreement at those distances where the extraction in the presence of explicit back-reaction is reliable. We thus reconfirm the results of \cite{Laine:2007qy} for the imaginary part, observing no significant effect of the back-reaction on the physics of scattering between static sources and medium constituents in the value of ${\rm Im}[V]$.

Interestingly, the values of the imaginary part, we obtain in lattice simulations, are significantly larger than those predicted by lattice perturbation theory (compare to  \cref{fig:NoSrcPot}). On the other hand, the Debye mass, governing the behavior of the real-part comes out at a very similar value. It would be interesting to explore, whether the agreement for ${\rm Im}[V]$ improves significantly when extending the perturbative calculation of Ref.~\cite{Laine:2007qy} to one higher order.

\section{Conclusion}
\label{sec:conclusion}

Classical statistical simulations of gauge fields in the presence of static sources provide us with qualitative insight into the dynamical binding mechanisms of heavy quarks under the strong interactions. In this paper, we have revisited how the presence of static sources affects the classical equations of motion, leading to an explicit source term in Gauss's Law, which was absent in previous works.

In order to scrutinize the binding of color sources, we extract the static potential acting between them, based on the state-of-the-art approach of computing the Wilson loop spectral function and carrying out a skewed Breit-Wigner fit to its lowest lying structure. In the absence of back-reaction this method allows us to confirm the results of a finite imaginary part first presented in \cite{Laine:2007qy} and extend them to larger spatial separation distances and a broader temperature range.  

By employing the proper Gauss's Law in the presence of static sources, we observe significant changes in the dynamics of the Wilson loop already on a qualitative level. Its values become complex and show oscillatory behavior which directly translate into finite values of the real-part of the static potential. We investigate its behavior at $\beta=20$ and find that it is well described by a Debye screened form. The imaginary part of the potential interestingly appears not to be affected by the inclusion of the back-reaction when compared to a simulation where the explicit back-reaction through the Gauss's law was absent.

The Wilson loop in conventional gauge theory embodies a reweighting from a theory without sources to the theory of interest with sources present. In the classical statistical theory it does not play this role and one may think about whether it is the most appropriate quantity to extract the interaction potential there. One argument to consider is that in contrast to $SU(3)$ gauge theory deployed in this study, we could just as well take $SU(2)$. In the fully quantum theory both theories are known to lead to interactions between color sources. However since the Wilson loop in $SU(2)$ must be purely real, it being the trace over the product of link variables, it cannot encode a finite real part of the potential, in the form observed here.

An alternative approach to determining the interaction potential has been put forward in the context of Euclidean lattice gauge theory in Ref.\cite{yanagihara_distribution_2019}. The authors propose to compute the expectation values of the gauge invariant energy momentum tensor and to inspect its spatial components. This stress tensor encodes the force acting locally on the faces of a given volume element. Integrating the net force up, allows one to compute the underlying potential. While this approach has been implemented with success in the Euclidean theory, the fact that the current discretizations of the energy momentum tensor are not conserving in Minkowski time prevent us from utilizing it straight away. It actually requires the development of a new conserving discretization scheme, which is currently work in progress.

Having elucidated the physics of static sources in the presence of classical statistical gauge fields in this study, we set out to compute the interaction between very heavy but still dynamical quarks in classical statistical lattice gauge theory in the future using a real-time implementation of the effective field theory NRQCD (for preliminary work in this direction see \cite{Lehmann:2020kjg}).

\section*{Acknowledgements}
The authors thank Y. Akamatsu for valuable and insightful discussions. A.L. was supported in full by the German Research Foundation (DFG) funded Collaborative Research Centre “SFB 1225 (ISOQUANT)”. A.R. gladly acknowledges support by the Research Council of Norway under the FRIPRO Young Research Talent grant 286883. The numerical simulations have been carried out on computing resources provided by  
UNINETT Sigma2 - the National Infrastructure for High Performance Computing and Data Storage in Norway under project NN9578K-QCDrtX "Real-time dynamics of nuclear matter under extreme conditions"  

\begin{appendix}

\section{Supplemental figures}
\label{sec:AppSupplementNum}

In order for the reader to judge the fidelity of our numerical simulations, we present here representative plots concerning the thermalization procedure, the conservation of the Gauss's law as well as conservation of energy. The first three figures are based on $\beta=4$ simulations in the absence of explicit sources from the Gauss's law. The remaining ones refer to the $\beta=20$ simulations in the presence of sources.

\begin{figure}
\includegraphics[scale=0.5]{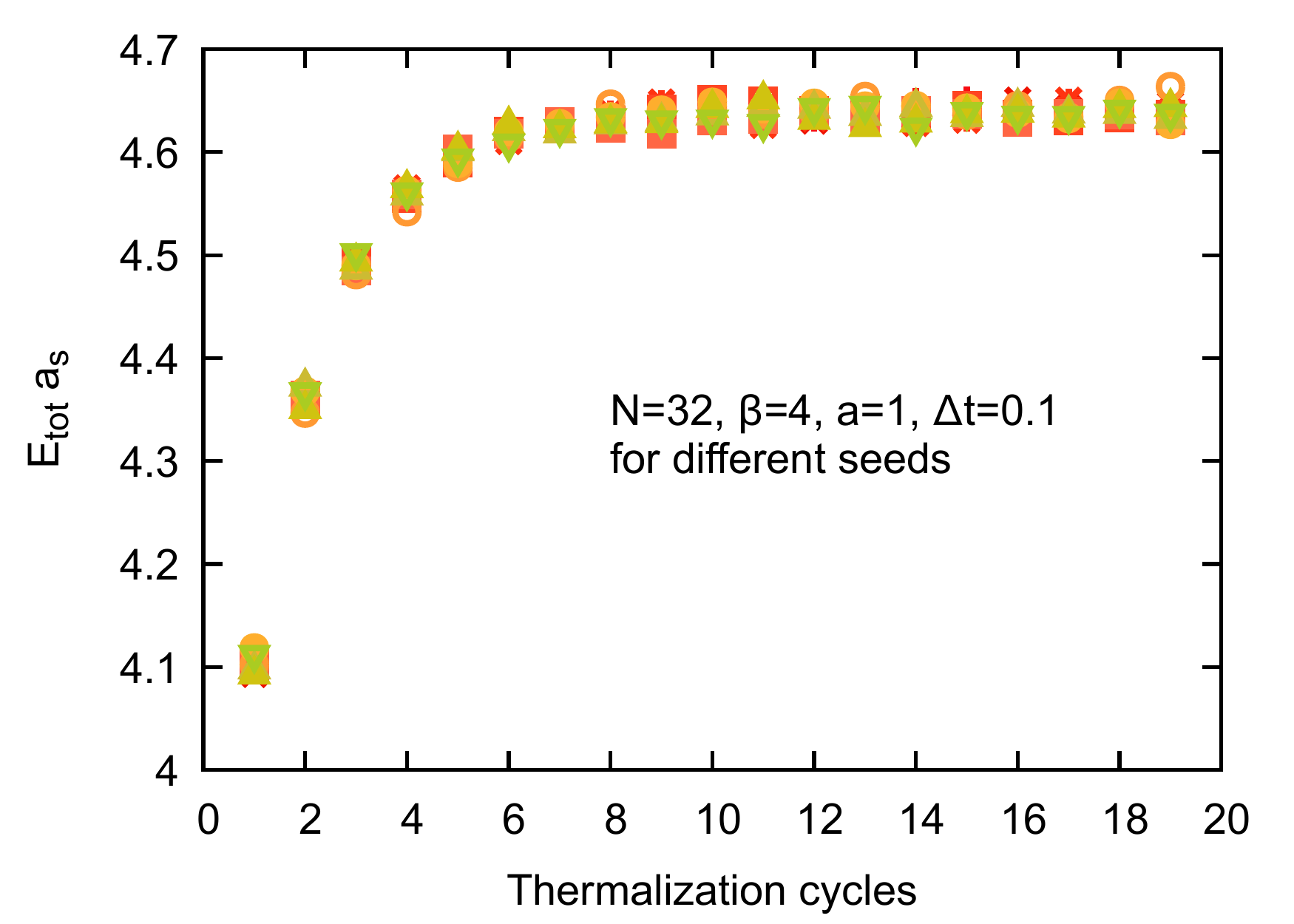}
\caption{Thermalization of the initial conditions in the absence of sources at $\beta=4$.}\label{fig:NoSrcThrm}
\end{figure}

Constructing the thermal initial conditions constitutes a major numerical cost in the classical statistical simulation of gauge fields. As shown in \cref{fig:NoSrcThrm} the total energy of the electric fields and spatial links after multiple mutual equilibration cycles approaches an asymptotic value, characteristic for that ensemble of thermal initial conditions. In order to keep the cost as low as possible we have chosen to draw and project the initial electric fields $N_{\rm init}=20$ times, after which the asymptotic state appears well realized.

\begin{figure}
\includegraphics[scale=0.5]{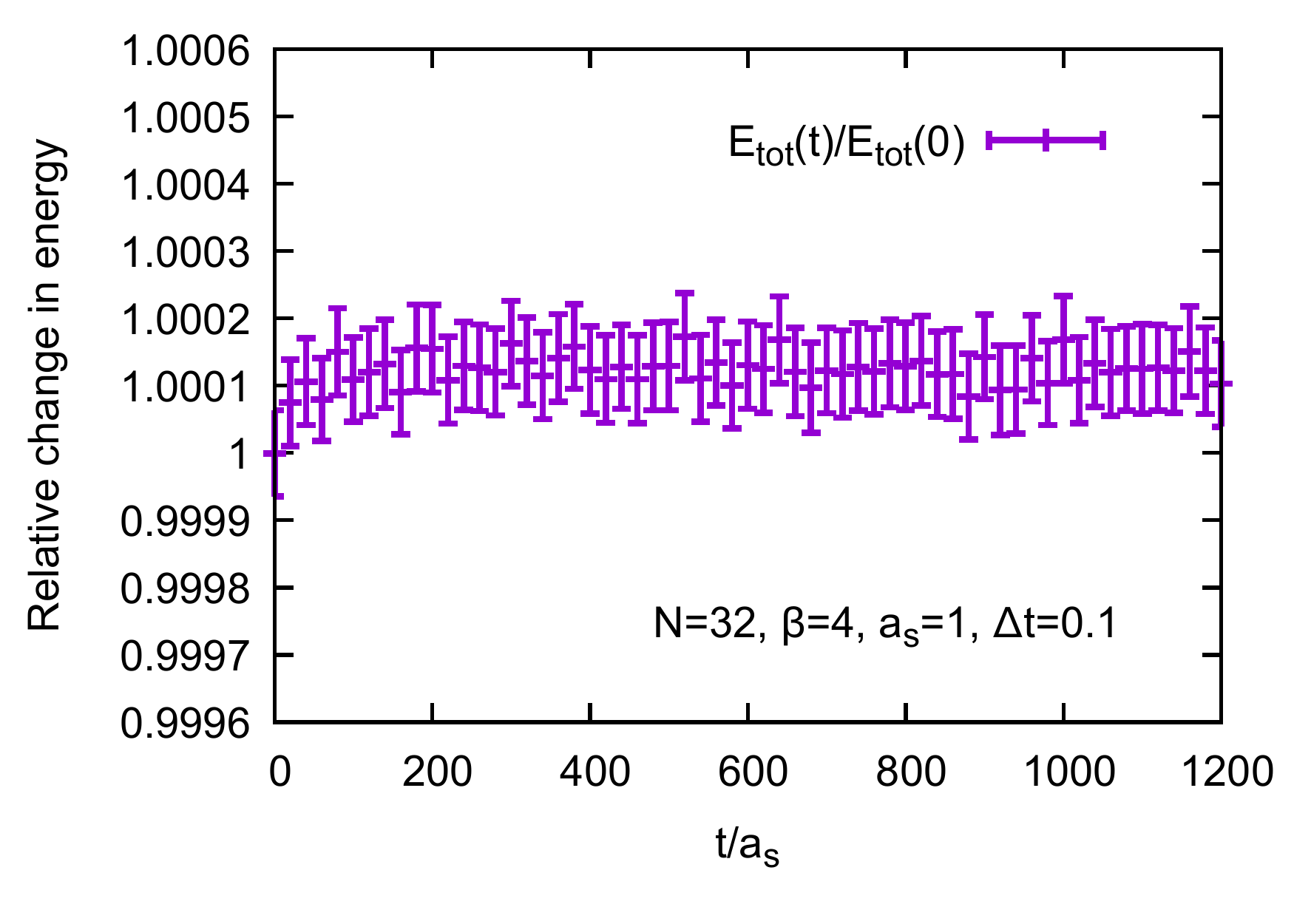}
\includegraphics[scale=0.5]{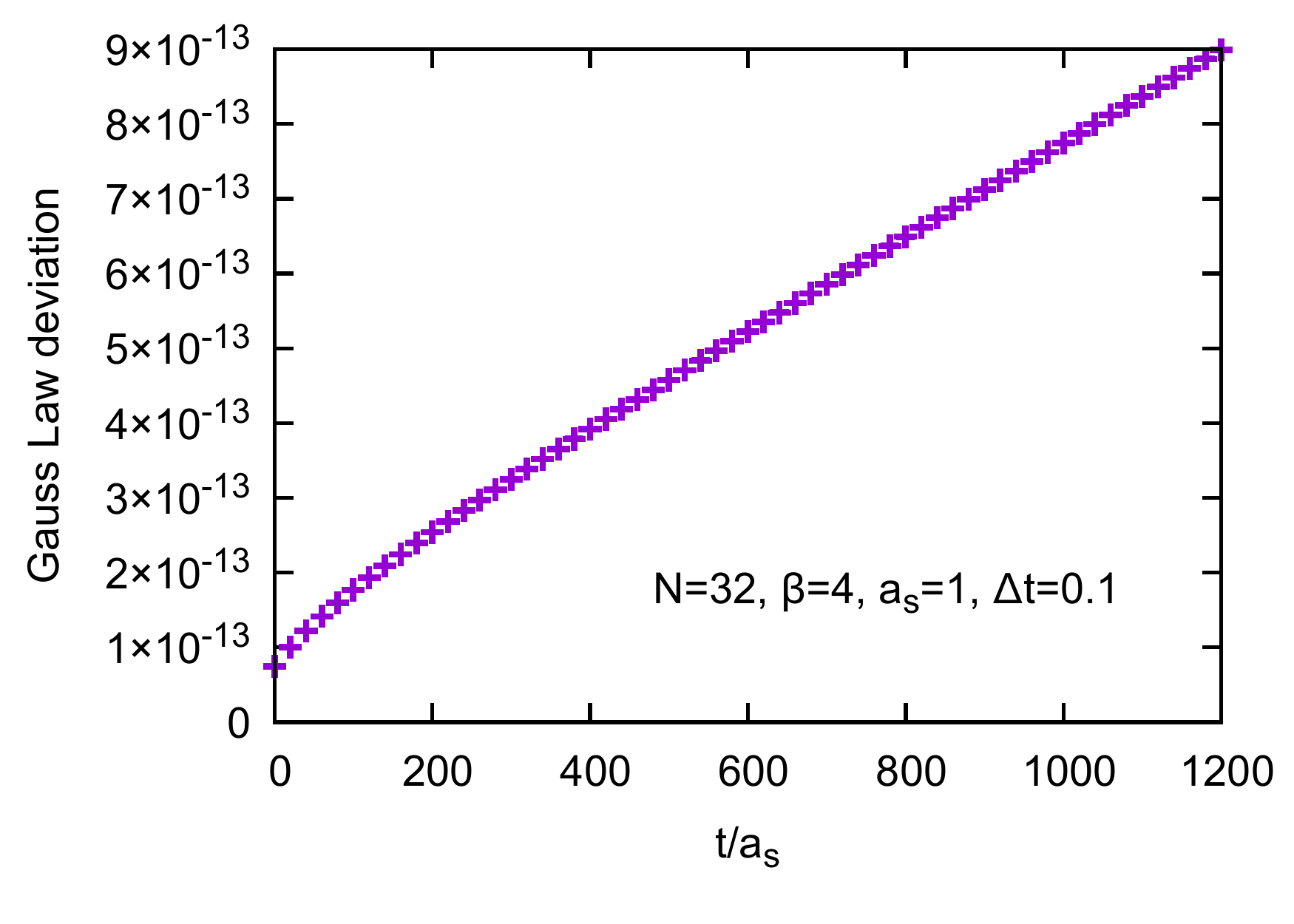}
\caption{Energy conservation (top) and Gauss's Law preservation (bottom) in the absence of sources at $\beta=4$.}\label{fig:NoSrcNrgGaussLaw}
\end{figure}

And indeed, when we observe the relative change in the total energy over the whole dynamical simulation, as plotted in the top panel of \cref{fig:NoSrcNrgGaussLaw}, we find that its value remains constant down to the sub-permille level. This in turn both tells us that thermal equilibrium between electric fields and links was reached and that the leap-frog time-stepping as symplectic solver realizes its potential to preserve energy in the discretized e.o.m..

The last check to be made in classical statistical simulations of Yang-Mills fields is whether the discretized dynamics preserve the Gauss's law. And while the time stepping introduces a finite deviation as shown in the bottom panel of \cref{fig:NoSrcNrgGaussLaw}, it remains fully insignificant over the whole time period of the simulation. Note that the starting value of around $5\times 10^{-13}$ arises from our choice of tolerance in projecting the electric fields to their physical values.

\begin{figure}
\includegraphics[scale=0.5]{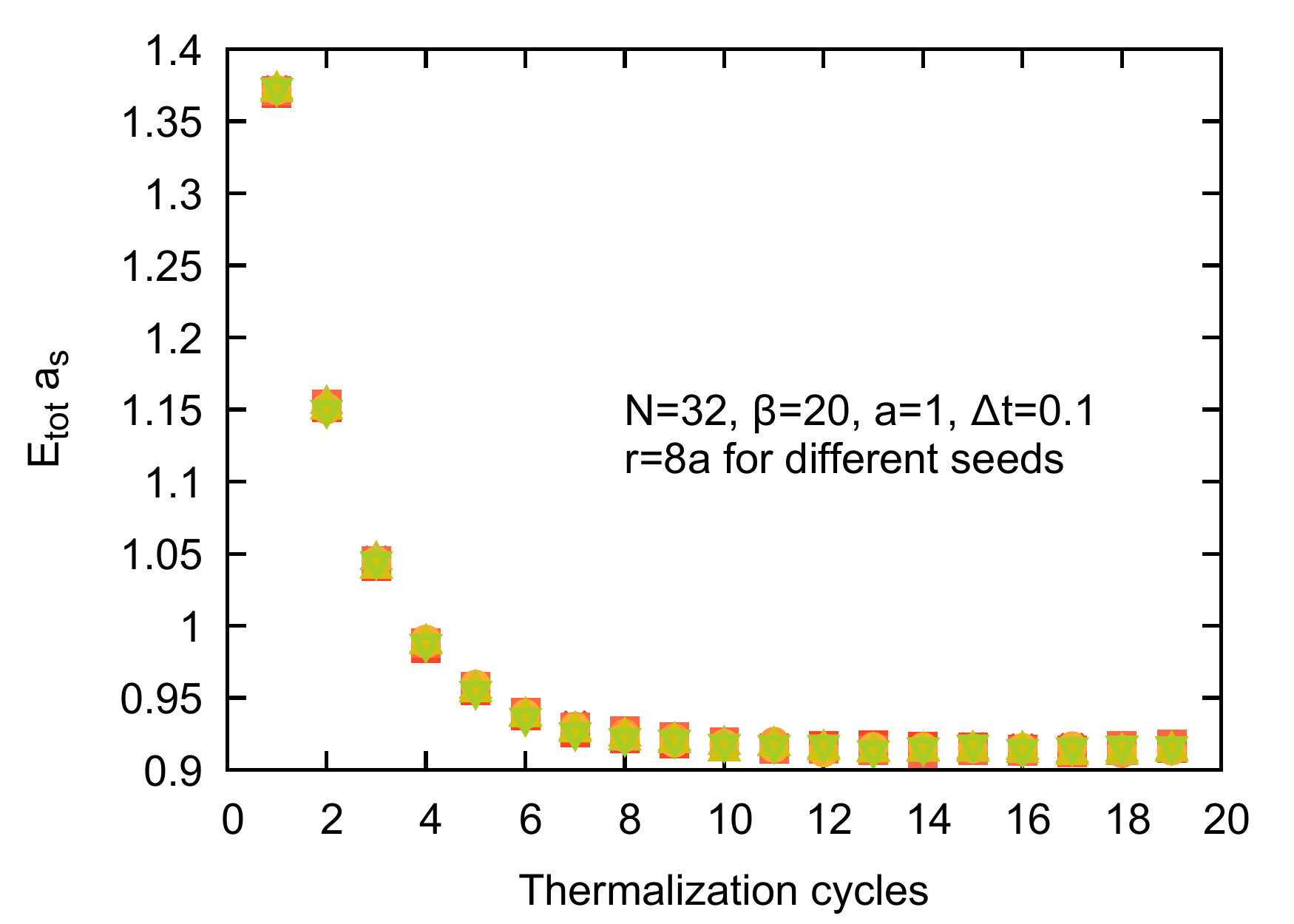}
\caption{Thermalization of the initial conditions in the presence of sources at spatial distance r=a at $\beta=20$.}\label{fig:SrcTherm}
\end{figure}

The simulations in the presence of explicit sources show similarly robust behavior. In \cref{fig:SrcTherm} we show the change in total energy over the $N_{\rm init}=20$ thermalization cycles after which the asymptotic value is well established.

\begin{figure}
\includegraphics[scale=0.5]{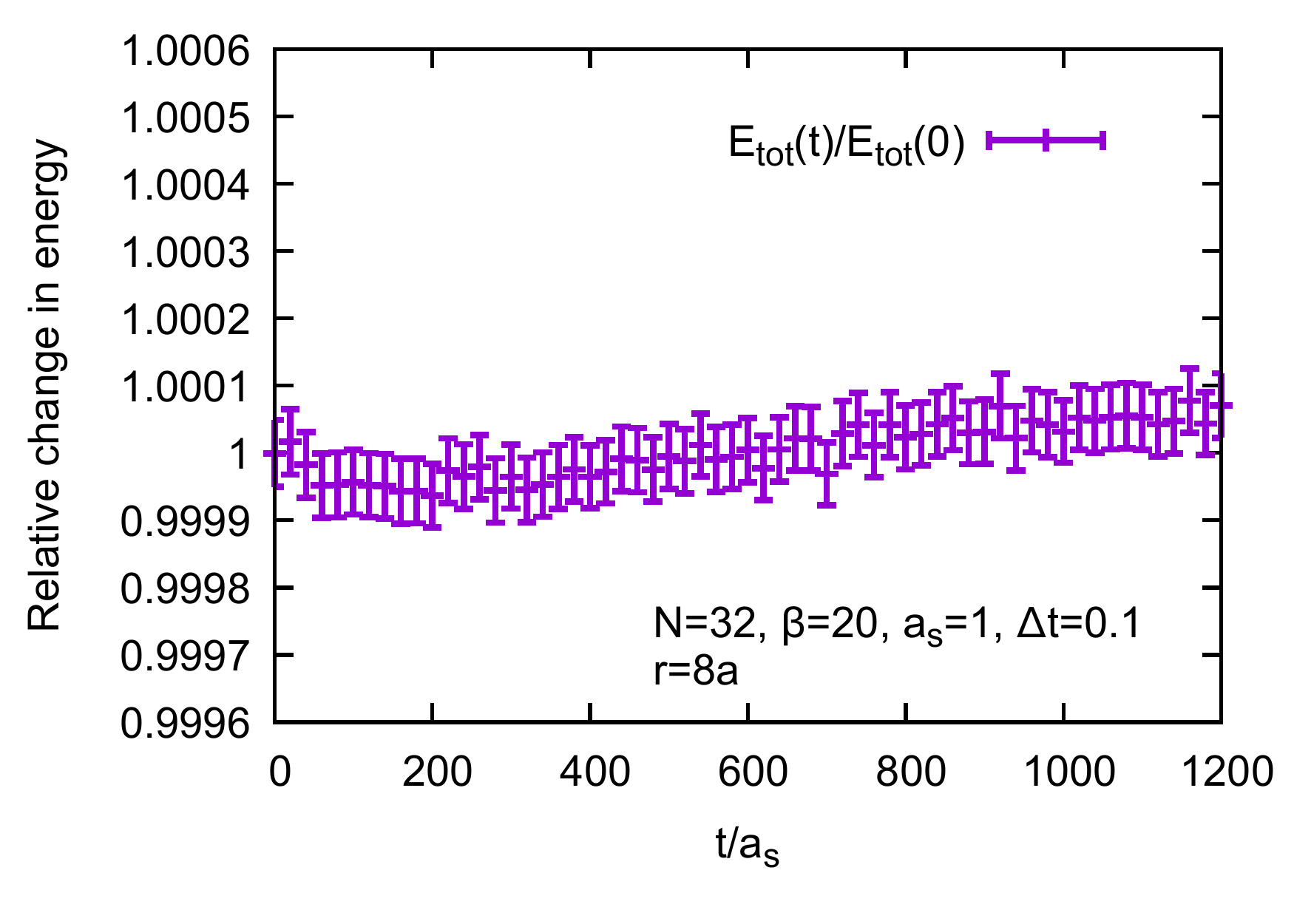}
\includegraphics[scale=0.5]{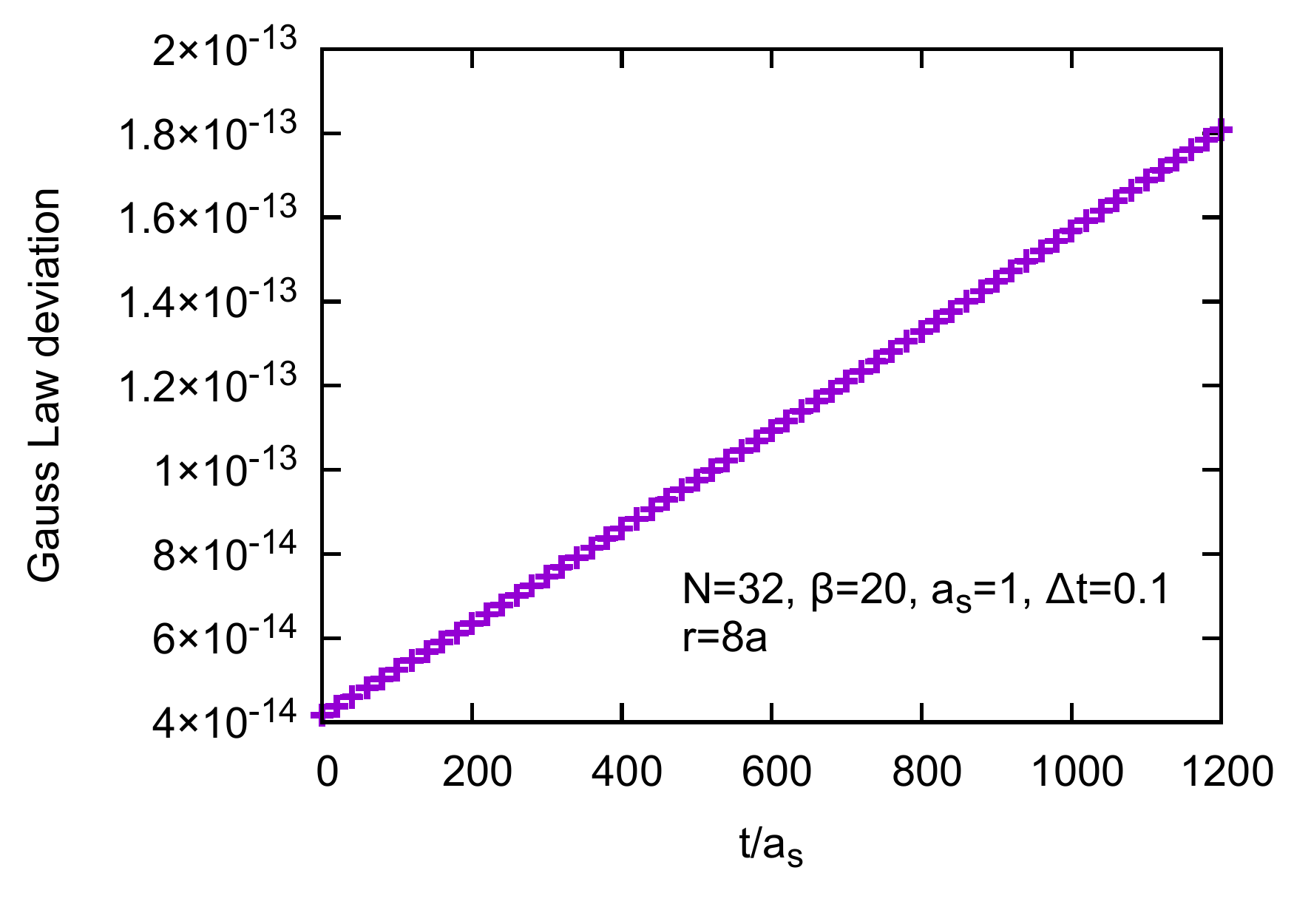}
\caption{Energy conservation (top) and Gauss's Law preservation (bottom) in the presence of sources at spatial distance r=a at $\beta=20$.}\label{fig:SrcNRGGauss}
\end{figure}

Taking a look at the evolution of the relative change in total energy during the actual simulation in the top panel of \cref{fig:SrcNRGGauss}, we find again that any residual change is on a below the per-mille level and thus as insignificant as in the case without sources.

The proper Gauss's law, the vital new ingredient in the determination of the proper interaction potential between static sources in lattice gauge theory, also turns out to be well preserved over the whole simulation time. As shown in the bottom panel of \cref{fig:SrcNRGGauss} it remains below $10^{-13}$ at all evolution steps. Again the start value of around $4\times 10^{-14}$ is related to the tolerance of projecting the initial electric fields to their physical subspace.

\end{appendix}  


\FloatBarrier

\begin{backmatter}

\section*{Competing interests}
  The authors declare that they have no competing interests.

\section*{Author's contributions}
    \begin{itemize}
        \item A. Lehmann: implementation real-time simulation, partial data analysis, writing
        \item A. Rothkopf: conceptualization, funding acquisition, supervision, independent implementation of the real-time simulation, data analysis, writing
    \end{itemize}


\bibliographystyle{stavanger-mathphys}


\bibliography{ThermalPotClassical}


\begin{thebibliography}{71}
\ifx \bisbn   \undefined \def \bisbn  #1{ISBN #1}\fi
\ifx \binits  \undefined \def \binits#1{#1}\fi
\ifx \bauthor  \undefined \def \bauthor#1{#1}\fi
\ifx \batitle  \undefined \def \batitle#1{#1}\fi
\ifx \bjtitle  \undefined \def \bjtitle#1{#1}\fi
\ifx \bvolume  \undefined \def \bvolume#1{\textbf{#1}}\fi
\ifx \byear  \undefined \def \byear#1{#1}\fi
\ifx \bissue  \undefined \def \bissue#1{#1}\fi
\ifx \bfpage  \undefined \def \bfpage#1{#1}\fi
\ifx \blpage  \undefined \def \blpage #1{#1}\fi
\ifx \burl  \undefined \def \burl#1{\textsf{#1}}\fi
\ifx \doiurl  \undefined \def \doiurl#1{\textsf{#1}}\fi
\ifx \betal  \undefined \def \betal{\textit{et al.}}\fi
\ifx \binstitute  \undefined \def \binstitute#1{#1}\fi
\ifx \binstitutionaled  \undefined \def \binstitutionaled#1{#1}\fi
\ifx \bctitle  \undefined \def \bctitle#1{#1}\fi
\ifx \beditor  \undefined \def \beditor#1{#1}\fi
\ifx \bpublisher  \undefined \def \bpublisher#1{#1}\fi
\ifx \bbtitle  \undefined \def \bbtitle#1{#1}\fi
\ifx \bedition  \undefined \def \bedition#1{#1}\fi
\ifx \bseriesno  \undefined \def \bseriesno#1{#1}\fi
\ifx \blocation  \undefined \def \blocation#1{#1}\fi
\ifx \bsertitle  \undefined \def \bsertitle#1{#1}\fi
\ifx \bsnm \undefined \def \bsnm#1{#1}\fi
\ifx \bsuffix \undefined \def \bsuffix#1{#1}\fi
\ifx \bparticle \undefined \def \bparticle#1{#1}\fi
\ifx \barticle \undefined \def \barticle#1{#1}\fi
\ifx \bconfdate \undefined \def \bconfdate #1{#1}\fi
\ifx \botherref \undefined \def \botherref #1{#1}\fi
\ifx \url \undefined \def \url#1{\textsf{#1}}\fi
\ifx \bchapter \undefined \def \bchapter#1{#1}\fi
\ifx \bbook \undefined \def \bbook#1{#1}\fi
\ifx \bcomment \undefined \def \bcomment#1{#1}\fi
\ifx \oauthor \undefined \def \oauthor#1{#1}\fi
\ifx \citeauthoryear \undefined \def \citeauthoryear#1{#1}\fi
\ifx \endbibitem  \undefined \def \endbibitem {}\fi
\ifx \bconflocation  \undefined \def \bconflocation#1{#1}\fi
\ifx \arxivurl  \undefined \def \arxivurl#1{\textsf{#1}}\fi
\csname PreBibitemsHook\endcsname

\bibitem{Brambilla:2010cs}
\begin{barticle}
\bauthor{\bsnm{Brambilla}, \binits{N.}}, \betal:
\batitle{{Heavy Quarkonium: Progress, Puzzles, and Opportunities}}.
\bjtitle{Eur. Phys. J.}
\bvolume{C71},
\bfpage{1534}
(\byear{2011}).
doi:\doiurl{10.1140/epjc/s10052-010-1534-9}.
\arxivurl{1010.5827}
\end{barticle}
\endbibitem

\bibitem{Rothkopf:2019ipj}
\begin{barticle}
\bauthor{\bsnm{Rothkopf}, \binits{A.}}:
\batitle{{Heavy Quarkonium in Extreme Conditions}}.
\bjtitle{Phys. Rept.}
\bvolume{858},
\bfpage{1}--\blpage{117}
(\byear{2020}).
doi:\doiurl{10.1016/j.physrep.2020.02.006}.
\arxivurl{1912.02253}
\end{barticle}
\endbibitem

\bibitem{Strickland:2020hyg}
\begin{barticle}
\bauthor{\bsnm{{Strickland, Michael}}}:
\batitle{{Pseudothermalization of the quark-gluon plasma}}.
\bjtitle{J. Phys. Conf. Ser.}
\bvolume{1602}(\bissue{1}),
\bfpage{012018}
(\byear{2020}).
doi:\doiurl{10.1088/1742-6596/1602/1/012018}.
\arxivurl{2006.09284}
\end{barticle}
\endbibitem

\bibitem{Berges:2020fwq}
\begin{botherref}
\oauthor{\bsnm{Berges}, \binits{J.}},
\oauthor{\bsnm{Heller}, \binits{M.P.}},
\oauthor{\bsnm{Mazeliauskas}, \binits{A.}},
\oauthor{\bsnm{Venugopalan}, \binits{R.}}:
{Thermalization in QCD: theoretical approaches, phenomenological applications,
  and interdisciplinary connections}
(2020).
\arxivurl{2005.12299}
\end{botherref}
\endbibitem

\bibitem{Shen:2020gef}
\begin{bchapter}
\bauthor{\bsnm{Shen}, \binits{C.}}:
\bctitle{{Studying QGP with flow: A theory overview}}.
In: \bbtitle{28th International Conference on Ultrarelativistic Nucleus-Nucleus
  Collisions}
(\byear{2020}).
\arxivurl{2001.11858}
\end{bchapter}
\endbibitem

\bibitem{Akamatsu:2020lej}
\begin{bchapter}
\bauthor{\bsnm{Akamatsu}, \binits{Y.}}:
\bctitle{{Approach to thermalization and hydrodynamics}}.
In: \bbtitle{28th International Conference on Ultrarelativistic Nucleus-Nucleus
  Collisions}
(\byear{2020}).
\arxivurl{2001.01429}
\end{bchapter}
\endbibitem

\bibitem{Kurkela:2019kdu}
\begin{barticle}
\bauthor{\bsnm{Kurkela}, \binits{A.}},
\bauthor{\bsnm{Mazeliauskas}, \binits{A.}},
\bauthor{\bsnm{Paquet}, \binits{J.-F.}},
\bauthor{\bsnm{Schlichting}, \binits{S.}},
\bauthor{\bsnm{Teaney}, \binits{D.}}:
\batitle{{Matching the non-equilibrium initial stage of heavy ion collisions to
  hydrodynamics with QCD kinetic theory}}.
\bjtitle{PoS}
\bvolume{Confinement2018},
\bfpage{152}
(\byear{2019}).
doi:\doiurl{10.22323/1.336.0152}
\end{barticle}
\endbibitem

\bibitem{Larsen:2019zqv}
\begin{barticle}
\bauthor{\bsnm{Larsen}, \binits{R.}},
\bauthor{\bsnm{Meinel}, \binits{S.}},
\bauthor{\bsnm{Mukherjee}, \binits{S.}},
\bauthor{\bsnm{Petreczky}, \binits{P.}}:
\batitle{{Excited bottomonia in quark-gluon plasma from lattice QCD}}.
\bjtitle{Phys. Lett. B}
\bvolume{800},
\bfpage{135119}
(\byear{2020}).
doi:\doiurl{10.1016/j.physletb.2019.135119}.
\arxivurl{1910.07374}
\end{barticle}
\endbibitem

\bibitem{Offler:2019eij}
\begin{barticle}
\bauthor{\bsnm{Offler}, \binits{S.}},
\bauthor{\bsnm{Aarts}, \binits{G.}},
\bauthor{\bsnm{Allton}, \binits{C.}},
\bauthor{\bsnm{Glesaaen}, \binits{J.}},
\bauthor{\bsnm{J\"ager}, \binits{B.}},
\bauthor{\bsnm{Kim}, \binits{S.}},
\bauthor{\bsnm{Lombardo}, \binits{M.P.}},
\bauthor{\bsnm{Ryan}, \binits{S.M.}},
\bauthor{\bsnm{Skullerud}, \binits{J.-I.}}:
\batitle{{News from bottomonium spectral functions in thermal QCD}}.
\bjtitle{PoS}
\bvolume{LATTICE2019},
\bfpage{076}
(\byear{2019}).
doi:\doiurl{10.22323/1.363.0076}.
\arxivurl{1912.12900}
\end{barticle}
\endbibitem

\bibitem{Larsen:2019bwy}
\begin{barticle}
\bauthor{\bsnm{Larsen}, \binits{R.}},
\bauthor{\bsnm{Meinel}, \binits{S.}},
\bauthor{\bsnm{Mukherjee}, \binits{S.}},
\bauthor{\bsnm{Petreczky}, \binits{P.}}:
\batitle{{Thermal broadening of bottomonia: Lattice nonrelativistic QCD with
  extended operators}}.
\bjtitle{Phys. Rev. D}
\bvolume{100}(\bissue{7}),
\bfpage{074506}
(\byear{2019}).
doi:\doiurl{10.1103/PhysRevD.100.074506}.
\arxivurl{1908.08437}
\end{barticle}
\endbibitem

\bibitem{Kim:2018yhk}
\begin{barticle}
\bauthor{\bsnm{Kim}, \binits{S.}},
\bauthor{\bsnm{Petreczky}, \binits{P.}},
\bauthor{\bsnm{Rothkopf}, \binits{A.}}:
\batitle{{Quarkonium in-medium properties from realistic lattice NRQCD}}.
\bjtitle{JHEP}
\bvolume{11},
\bfpage{088}
(\byear{2018}).
doi:\doiurl{10.1007/JHEP11(2018)088}.
\arxivurl{1808.08781}
\end{barticle}
\endbibitem

\bibitem{Burnier:2017bod}
\begin{barticle}
\bauthor{\bsnm{Burnier}, \binits{Y.}},
\bauthor{\bsnm{Ding}, \binits{H.-T.}},
\bauthor{\bsnm{Kaczmarek}, \binits{O.}},
\bauthor{\bsnm{Kruse}, \binits{A.-L.}},
\bauthor{\bsnm{Laine}, \binits{M.}},
\bauthor{\bsnm{Ohno}, \binits{H.}},
\bauthor{\bsnm{Sandmeyer}, \binits{H.}}:
\batitle{{Thermal quarkonium physics in the pseudoscalar channel}}.
\bjtitle{JHEP}
\bvolume{11},
\bfpage{206}
(\byear{2017}).
doi:\doiurl{10.1007/JHEP11(2017)206}.
\arxivurl{1709.07612}
\end{barticle}
\endbibitem

\bibitem{Burnier:2015tda}
\begin{barticle}
\bauthor{\bsnm{Burnier}, \binits{Y.}},
\bauthor{\bsnm{Kaczmarek}, \binits{O.}},
\bauthor{\bsnm{Rothkopf}, \binits{A.}}:
\batitle{{Quarkonium at finite temperature: Towards realistic phenomenology
  from first principles}}.
\bjtitle{JHEP}
\bvolume{12},
\bfpage{101}
(\byear{2015}).
doi:\doiurl{10.1007/JHEP12(2015)101}.
\arxivurl{1509.07366}
\end{barticle}
\endbibitem

\bibitem{Burnier:2016kqm}
\begin{barticle}
\bauthor{\bsnm{Burnier}, \binits{Y.}},
\bauthor{\bsnm{Kaczmarek}, \binits{O.}},
\bauthor{\bsnm{Rothkopf}, \binits{A.}}:
\batitle{{In-medium P-wave quarkonium from the complex lattice QCD potential}}.
\bjtitle{JHEP}
\bvolume{10},
\bfpage{032}
(\byear{2016}).
doi:\doiurl{10.1007/JHEP10(2016)032}.
\arxivurl{1606.06211}
\end{barticle}
\endbibitem

\bibitem{Song:2020kka}
\begin{botherref}
\oauthor{\bsnm{Song}, \binits{T.}},
\oauthor{\bsnm{Gubler}, \binits{P.}},
\oauthor{\bsnm{Hong}, \binits{J.}},
\oauthor{\bsnm{Lee}, \binits{S.H.}},
\oauthor{\bsnm{Morita}, \binits{K.}}:
{$J/\psi$ near $T_c$}
(2020).
\arxivurl{2009.08741}
\end{botherref}
\endbibitem

\bibitem{Braga:2017bml}
\begin{barticle}
\bauthor{\bsnm{Braga}, \binits{N.R.F.}},
\bauthor{\bsnm{Ferreira}, \binits{L.F.}},
\bauthor{\bsnm{Vega}, \binits{A.}}:
\batitle{{Holographic model for charmonium dissociation}}.
\bjtitle{Phys. Lett.}
\bvolume{B774},
\bfpage{476}--\blpage{481}
(\byear{2017}).
doi:\doiurl{10.1016/j.physletb.2017.10.013}.
\arxivurl{1709.05326}
\end{barticle}
\endbibitem

\bibitem{Grigoryan:2010pj}
\begin{barticle}
\bauthor{\bsnm{Grigoryan}, \binits{H.R.}},
\bauthor{\bsnm{Hohler}, \binits{P.M.}},
\bauthor{\bsnm{Stephanov}, \binits{M.A.}}:
\batitle{{Towards the Gravity Dual of Quarkonium in the Strongly Coupled QCD
  Plasma}}.
\bjtitle{Phys. Rev.}
\bvolume{D82},
\bfpage{026005}
(\byear{2010}).
doi:\doiurl{10.1103/PhysRevD.82.026005}.
\arxivurl{1003.1138}
\end{barticle}
\endbibitem

\bibitem{Fujita:2009ca}
\begin{barticle}
\bauthor{\bsnm{Fujita}, \binits{M.}},
\bauthor{\bsnm{Kikuchi}, \binits{T.}},
\bauthor{\bsnm{Fukushima}, \binits{K.}},
\bauthor{\bsnm{Misumi}, \binits{T.}},
\bauthor{\bsnm{Murata}, \binits{M.}}:
\batitle{{Melting Spectral Functions of the Scalar and Vector Mesons in a
  Holographic QCD Model}}.
\bjtitle{Phys. Rev.}
\bvolume{D81},
\bfpage{065024}
(\byear{2010}).
doi:\doiurl{10.1103/PhysRevD.81.065024}.
\arxivurl{0911.2298}
\end{barticle}
\endbibitem

\bibitem{Borghini:2011ms}
\begin{barticle}
\bauthor{\bsnm{Borghini}, \binits{N.}},
\bauthor{\bsnm{Gombeaud}, \binits{C.}}:
\batitle{{Heavy quarkonia in a medium as a quantum dissipative system: Master
  equation approach}}.
\bjtitle{Eur. Phys. J.}
\bvolume{C72},
\bfpage{2000}
(\byear{2012}).
doi:\doiurl{10.1140/epjc/s10052-012-2000-7}.
\arxivurl{1109.4271}
\end{barticle}
\endbibitem

\bibitem{Akamatsu:2011se}
\begin{barticle}
\bauthor{\bsnm{Akamatsu}, \binits{Y.}},
\bauthor{\bsnm{Rothkopf}, \binits{A.}}:
\batitle{{Stochastic potential and quantum decoherence of heavy quarkonium in
  the quark-gluon plasma}}.
\bjtitle{Phys. Rev.}
\bvolume{D85},
\bfpage{105011}
(\byear{2012}).
doi:\doiurl{10.1103/PhysRevD.85.105011}.
\arxivurl{1110.1203}
\end{barticle}
\endbibitem

\bibitem{Akamatsu:2012vt}
\begin{barticle}
\bauthor{\bsnm{Akamatsu}, \binits{Y.}}:
\batitle{{Real-time quantum dynamics of heavy quark systems at high
  temperature}}.
\bjtitle{Phys. Rev.}
\bvolume{D87}(\bissue{4}),
\bfpage{045016}
(\byear{2013}).
doi:\doiurl{10.1103/PhysRevD.87.045016}.
\arxivurl{1209.5068}
\end{barticle}
\endbibitem

\bibitem{Brambilla:2020qwo}
\begin{botherref}
\oauthor{\bsnm{Brambilla}, \binits{N.}},
\oauthor{\bsnm{Escobedo}, \binits{M.A.}},
\oauthor{\bsnm{Strickland}, \binits{M.}},
\oauthor{\bsnm{Vairo}, \binits{A.}},
\oauthor{\bsnm{Vander~Griend}, \binits{P.}},
\oauthor{\bsnm{Weber}, \binits{J.H.}}:
{Bottomonium suppression in an open quantum system using the quantum
  trajectories method}
(2020).
\arxivurl{2012.01240}
\end{botherref}
\endbibitem

\bibitem{deJong:2020tvx}
\begin{botherref}
\oauthor{\bsnm{De~Jong}, \binits{W.A.}},
\oauthor{\bsnm{Metcalf}, \binits{M.}},
\oauthor{\bsnm{Mulligan}, \binits{J.}},
\oauthor{\bsnm{P\l{}osko\'n}, \binits{M.}},
\oauthor{\bsnm{Ringer}, \binits{F.}},
\oauthor{\bsnm{Yao}, \binits{X.}}:
{Quantum simulation of open quantum systems in heavy-ion collisions}
(2020).
\arxivurl{2010.03571}
\end{botherref}
\endbibitem

\bibitem{Yao:2020eqy}
\begin{botherref}
\oauthor{\bsnm{Yao}, \binits{X.}},
\oauthor{\bsnm{Mehen}, \binits{T.}}:
{Quarkonium Semiclassical Transport in Quark-Gluon Plasma: Factorization and
  Quantum Correction}
(2020).
\arxivurl{2009.02408}
\end{botherref}
\endbibitem

\bibitem{Yao:2020xzw}
\begin{botherref}
\oauthor{\bsnm{Yao}, \binits{X.}},
\oauthor{\bsnm{Ke}, \binits{W.}},
\oauthor{\bsnm{Xu}, \binits{Y.}},
\oauthor{\bsnm{Bass}, \binits{S.A.}},
\oauthor{\bsnm{M\"uller}, \binits{B.}}:
{Coupled Boltzmann Transport Equations of Heavy Quarks and Quarkonia in
  Quark-Gluon Plasma}
(2020).
\arxivurl{2004.06746}
\end{botherref}
\endbibitem

\bibitem{Alund:2020ctu}
\begin{botherref}
\oauthor{\bsnm{\r{A}lund}, \binits{O.}},
\oauthor{\bsnm{Akamatsu}, \binits{Y.}},
\oauthor{\bsnm{Laur\'en}, \binits{F.}},
\oauthor{\bsnm{Miura}, \binits{T.}},
\oauthor{\bsnm{Nordstr\"om}, \binits{J.}},
\oauthor{\bsnm{Rothkopf}, \binits{A.}}:
{Trace preserving quantum dynamics using a novel reparametrization-neutral
  summation-by-parts difference operator}
(2020).
\arxivurl{2004.04406}
\end{botherref}
\endbibitem

\bibitem{Miura:2019ssi}
\begin{barticle}
\bauthor{\bsnm{Miura}, \binits{T.}},
\bauthor{\bsnm{Akamatsu}, \binits{Y.}},
\bauthor{\bsnm{Asakawa}, \binits{M.}},
\bauthor{\bsnm{Rothkopf}, \binits{A.}}:
\batitle{{Quantum Brownian motion of a heavy quark pair in the quark-gluon
  plasma}}.
\bjtitle{Phys. Rev. D}
\bvolume{101}(\bissue{3}),
\bfpage{034011}
(\byear{2020}).
doi:\doiurl{10.1103/PhysRevD.101.034011}.
\arxivurl{1908.06293}
\end{barticle}
\endbibitem

\bibitem{Brambilla:2019tpt}
\begin{barticle}
\bauthor{\bsnm{Brambilla}, \binits{N.}},
\bauthor{\bsnm{Escobedo}, \binits{M.A.}},
\bauthor{\bsnm{Vairo}, \binits{A.}},
\bauthor{\bsnm{Vander~Griend}, \binits{P.}}:
\batitle{{Transport coefficients from in medium quarkonium dynamics}}.
\bjtitle{Phys. Rev. D}
\bvolume{100}(\bissue{5}),
\bfpage{054025}
(\byear{2019}).
doi:\doiurl{10.1103/PhysRevD.100.054025}.
\arxivurl{1903.08063}
\end{barticle}
\endbibitem

\bibitem{Yao:2018sgn}
\begin{barticle}
\bauthor{\bsnm{Yao}, \binits{X.}},
\bauthor{\bsnm{Müller}, \binits{B.}}:
\batitle{{Quarkonium inside the quark-gluon plasma: Diffusion, dissociation,
  recombination, and energy loss}}.
\bjtitle{Phys. Rev.}
\bvolume{D100}(\bissue{1}),
\bfpage{014008}
(\byear{2019}).
doi:\doiurl{10.1103/PhysRevD.100.014008}.
\arxivurl{1811.09644}
\end{barticle}
\endbibitem

\bibitem{Blaizot:2018oev}
\begin{barticle}
\bauthor{\bsnm{Blaizot}, \binits{J.-P.}},
\bauthor{\bsnm{Escobedo}, \binits{M.A.}}:
\batitle{{Approach to equilibrium of a quarkonium in a quark-gluon plasma}}.
\bjtitle{Phys. Rev.}
\bvolume{D98}(\bissue{7}),
\bfpage{074007}
(\byear{2018}).
doi:\doiurl{10.1103/PhysRevD.98.074007}.
\arxivurl{1803.07996}
\end{barticle}
\endbibitem

\bibitem{Brambilla:2017zei}
\begin{barticle}
\bauthor{\bsnm{Brambilla}, \binits{N.}},
\bauthor{\bsnm{Escobedo}, \binits{M.A.}},
\bauthor{\bsnm{Soto}, \binits{J.}},
\bauthor{\bsnm{Vairo}, \binits{A.}}:
\batitle{{Heavy quarkonium suppression in a fireball}}.
\bjtitle{Phys. Rev. D}
\bvolume{97}(\bissue{7}),
\bfpage{074009}
(\byear{2018}).
doi:\doiurl{10.1103/PhysRevD.97.074009}.
\arxivurl{1711.04515}
\end{barticle}
\endbibitem

\bibitem{Kajimoto:2017rel}
\begin{barticle}
\bauthor{\bsnm{Kajimoto}, \binits{S.}},
\bauthor{\bsnm{Akamatsu}, \binits{Y.}},
\bauthor{\bsnm{Asakawa}, \binits{M.}},
\bauthor{\bsnm{Rothkopf}, \binits{A.}}:
\batitle{{Dynamical dissociation of quarkonia by wave function decoherence}}.
\bjtitle{Phys. Rev. D}
\bvolume{97}(\bissue{1}),
\bfpage{014003}
(\byear{2018}).
doi:\doiurl{10.1103/PhysRevD.97.014003}.
\arxivurl{1705.03365}
\end{barticle}
\endbibitem

\bibitem{Yao:2017fuc}
\begin{barticle}
\bauthor{\bsnm{Yao}, \binits{X.}},
\bauthor{\bsnm{Müller}, \binits{B.}}:
\batitle{{Approach to equilibrium of quarkonium in quark-gluon plasma}}.
\bjtitle{Phys. Rev.}
\bvolume{C97}(\bissue{1}),
\bfpage{014908}
(\byear{2018}).
doi:\doiurl{10.1103/PhysRevC.97.049903, 10.1103/PhysRevC.97.014908}.
\bcomment{[Erratum: Phys. Rev.C97,no.4,049903(2018)]}.
\arxivurl{1709.03529}
\end{barticle}
\endbibitem

\bibitem{Akamatsu:2020ypb}
\begin{botherref}
\oauthor{\bsnm{Akamatsu}, \binits{Y.}}:
{Quarkonium in Quark-Gluon Plasma: Open Quantum System Approaches Re-examined}
(2020).
\arxivurl{2009.10559}
\end{botherref}
\endbibitem

\bibitem{Brambilla:2004jw}
\begin{barticle}
\bauthor{\bsnm{Brambilla}, \binits{N.}},
\bauthor{\bsnm{Pineda}, \binits{A.}},
\bauthor{\bsnm{Soto}, \binits{J.}},
\bauthor{\bsnm{Vairo}, \binits{A.}}:
\batitle{{Effective field theories for heavy quarkonium}}.
\bjtitle{Rev. Mod. Phys.}
\bvolume{77},
\bfpage{1423}
(\byear{2005}).
doi:\doiurl{10.1103/RevModPhys.77.1423}.
\arxivurl{hep-ph/0410047}
\end{barticle}
\endbibitem

\bibitem{Laine:2006ns}
\begin{barticle}
\bauthor{\bsnm{Laine}, \binits{M.}},
\bauthor{\bsnm{Philipsen}, \binits{O.}},
\bauthor{\bsnm{Romatschke}, \binits{P.}},
\bauthor{\bsnm{Tassler}, \binits{M.}}:
\batitle{{Real-time static potential in hot QCD}}.
\bjtitle{JHEP}
\bvolume{03},
\bfpage{054}
(\byear{2007}).
doi:\doiurl{10.1088/1126-6708/2007/03/054}.
\arxivurl{hep-ph/0611300}
\end{barticle}
\endbibitem

\bibitem{Beraudo:2007ky}
\begin{barticle}
\bauthor{\bsnm{Beraudo}, \binits{A.}},
\bauthor{\bsnm{Blaizot}, \binits{J.-P.}},
\bauthor{\bsnm{Ratti}, \binits{C.}}:
\batitle{{Real and imaginary-time Q anti-Q correlators in a thermal medium}}.
\bjtitle{Nucl. Phys.}
\bvolume{A806},
\bfpage{312}--\blpage{338}
(\byear{2008}).
doi:\doiurl{10.1016/j.nuclphysa.2008.03.001}.
\arxivurl{0712.4394}
\end{barticle}
\endbibitem

\bibitem{Brambilla:2008cx}
\begin{barticle}
\bauthor{\bsnm{Brambilla}, \binits{N.}},
\bauthor{\bsnm{Ghiglieri}, \binits{J.}},
\bauthor{\bsnm{Vairo}, \binits{A.}},
\bauthor{\bsnm{Petreczky}, \binits{P.}}:
\batitle{{Static quark-antiquark pairs at finite temperature}}.
\bjtitle{Phys. Rev.}
\bvolume{D78},
\bfpage{014017}
(\byear{2008}).
doi:\doiurl{10.1103/PhysRevD.78.014017}.
\arxivurl{0804.0993}
\end{barticle}
\endbibitem

\bibitem{Rothkopf:2011db}
\begin{barticle}
\bauthor{\bsnm{Rothkopf}, \binits{A.}},
\bauthor{\bsnm{Hatsuda}, \binits{T.}},
\bauthor{\bsnm{Sasaki}, \binits{S.}}:
\batitle{{Complex Heavy-Quark Potential at Finite Temperature from Lattice
  QCD}}.
\bjtitle{Phys. Rev. Lett.}
\bvolume{108},
\bfpage{162001}
(\byear{2012}).
doi:\doiurl{10.1103/PhysRevLett.108.162001}.
\arxivurl{1108.1579}
\end{barticle}
\endbibitem

\bibitem{Burnier:2014ssa}
\begin{barticle}
\bauthor{\bsnm{Burnier}, \binits{Y.}},
\bauthor{\bsnm{Kaczmarek}, \binits{O.}},
\bauthor{\bsnm{Rothkopf}, \binits{A.}}:
\batitle{{Static quark-antiquark potential in the quark-gluon plasma from
  lattice QCD}}.
\bjtitle{Phys. Rev. Lett.}
\bvolume{114}(\bissue{8}),
\bfpage{082001}
(\byear{2015}).
doi:\doiurl{10.1103/PhysRevLett.114.082001}.
\arxivurl{1410.2546}
\end{barticle}
\endbibitem

\bibitem{Burnier:2016mxc}
\begin{barticle}
\bauthor{\bsnm{Burnier}, \binits{Y.}},
\bauthor{\bsnm{Rothkopf}, \binits{A.}}:
\batitle{{Complex heavy-quark potential and Debye mass in a gluonic medium from
  lattice QCD}}.
\bjtitle{Phys. Rev. D}
\bvolume{95}(\bissue{5}),
\bfpage{054511}
(\byear{2017}).
doi:\doiurl{10.1103/PhysRevD.95.054511}.
\arxivurl{1607.04049}
\end{barticle}
\endbibitem

\bibitem{Petreczky:2017aiz}
\begin{barticle}
\bauthor{\bsnm{Petreczky}, \binits{P.}},
\bauthor{\bsnm{Weber}, \binits{J.}}:
\batitle{{Lattice Calculations of Heavy Quark Potential at Finite
  Temperature}}.
\bjtitle{Nucl. Phys. A}
\bvolume{967},
\bfpage{592}--\blpage{595}
(\byear{2017}).
doi:\doiurl{10.1016/j.nuclphysa.2017.04.011}.
\arxivurl{1704.08573}
\end{barticle}
\endbibitem

\bibitem{Petreczky:2018xuh}
\begin{barticle}
\bauthor{\bsnm{Petreczky}, \binits{P.}},
\bauthor{\bsnm{Rothkopf}, \binits{A.}},
\bauthor{\bsnm{Weber}, \binits{J.}}:
\batitle{{Realistic in-medium heavy-quark potential from high statistics
  lattice QCD simulations}}.
\bjtitle{Nucl. Phys. A}
\bvolume{982},
\bfpage{735}--\blpage{738}
(\byear{2019}).
doi:\doiurl{10.1016/j.nuclphysa.2018.10.012}.
\arxivurl{1810.02230}
\end{barticle}
\endbibitem

\bibitem{Lafferty:2019jpr}
\begin{barticle}
\bauthor{\bsnm{Lafferty}, \binits{D.}},
\bauthor{\bsnm{Rothkopf}, \binits{A.}}:
\batitle{{Improved Gauss law model and in-medium heavy quarkonium at finite
  density and velocity}}.
\bjtitle{Phys. Rev. D}
\bvolume{101}(\bissue{5}),
\bfpage{056010}
(\byear{2020}).
doi:\doiurl{10.1103/PhysRevD.101.056010}.
\arxivurl{1906.00035}
\end{barticle}
\endbibitem

\bibitem{Burnier:2012az}
\begin{barticle}
\bauthor{\bsnm{Burnier}, \binits{Y.}},
\bauthor{\bsnm{Rothkopf}, \binits{A.}}:
\batitle{{Disentangling the timescales behind the non-perturbative heavy quark
  potential}}.
\bjtitle{Phys. Rev. D}
\bvolume{86},
\bfpage{051503}
(\byear{2012}).
doi:\doiurl{10.1103/PhysRevD.86.051503}.
\arxivurl{1208.1899}
\end{barticle}
\endbibitem

\bibitem{Burnier:2013fca}
\begin{barticle}
\bauthor{\bsnm{Burnier}, \binits{Y.}},
\bauthor{\bsnm{Rothkopf}, \binits{A.}}:
\batitle{{A hard thermal loop benchmark for the extraction of the
  nonperturbative $Q\bar{Q}$ potential}}.
\bjtitle{Phys. Rev. D}
\bvolume{87},
\bfpage{114019}
(\byear{2013}).
doi:\doiurl{10.1103/PhysRevD.87.114019}.
\arxivurl{1304.4154}
\end{barticle}
\endbibitem

\bibitem{Rothkopf:2019dzu}
\begin{barticle}
\bauthor{\bsnm{Rothkopf}, \binits{A.}}:
\batitle{{Bayesian techniques and applications to QCD}}.
\bjtitle{PoS}
\bvolume{Confinement2018},
\bfpage{026}
(\byear{2018}).
doi:\doiurl{10.22323/1.336.0026}.
\arxivurl{1903.02293}
\end{barticle}
\endbibitem

\bibitem{DOnofrio:2014rug}
\begin{barticle}
\bauthor{\bsnm{D'Onofrio}, \binits{M.}},
\bauthor{\bsnm{Rummukainen}, \binits{K.}},
\bauthor{\bsnm{Tranberg}, \binits{A.}}:
\batitle{{Sphaleron Rate in the Minimal Standard Model}}.
\bjtitle{Phys. Rev. Lett.}
\bvolume{113}(\bissue{14}),
\bfpage{141602}
(\byear{2014}).
doi:\doiurl{10.1103/PhysRevLett.113.141602}.
\arxivurl{1404.3565}
\end{barticle}
\endbibitem

\bibitem{Micha:2004bv}
\begin{barticle}
\bauthor{\bsnm{Micha}, \binits{R.}},
\bauthor{\bsnm{Tkachev}, \binits{I.I.}}:
\batitle{{Turbulent thermalization}}.
\bjtitle{Phys. Rev. D}
\bvolume{70},
\bfpage{043538}
(\byear{2004}).
doi:\doiurl{10.1103/PhysRevD.70.043538}.
\arxivurl{hep-ph/0403101}
\end{barticle}
\endbibitem

\bibitem{Berges:2008wm}
\begin{barticle}
\bauthor{\bsnm{Berges}, \binits{J.}},
\bauthor{\bsnm{Rothkopf}, \binits{A.}},
\bauthor{\bsnm{Schmidt}, \binits{J.}}:
\batitle{{Non-thermal fixed points: Effective weak-coupling for strongly
  correlated systems far from equilibrium}}.
\bjtitle{Phys. Rev. Lett.}
\bvolume{101},
\bfpage{041603}
(\byear{2008}).
doi:\doiurl{10.1103/PhysRevLett.101.041603}.
\arxivurl{0803.0131}
\end{barticle}
\endbibitem

\bibitem{Berges:2016nru}
\begin{barticle}
\bauthor{\bsnm{Berges}, \binits{J.}},
\bauthor{\bsnm{Wallisch}, \binits{B.}}:
\batitle{{Nonthermal Fixed Points in Quantum Field Theory Beyond the
  Weak-Coupling Limit}}.
\bjtitle{Phys. Rev. D}
\bvolume{95}(\bissue{3}),
\bfpage{036016}
(\byear{2017}).
doi:\doiurl{10.1103/PhysRevD.95.036016}.
\arxivurl{1607.02160}
\end{barticle}
\endbibitem

\bibitem{Chatrchyan:2020syc}
\begin{botherref}
\oauthor{\bsnm{Chatrchyan}, \binits{A.}},
\oauthor{\bsnm{Geier}, \binits{K.T.}},
\oauthor{\bsnm{Oberthaler}, \binits{M.K.}},
\oauthor{\bsnm{Berges}, \binits{J.}},
\oauthor{\bsnm{Hauke}, \binits{P.}}:
{Analog reheating of the early universe in the laboratory}
(2020).
\arxivurl{2008.02290}
\end{botherref}
\endbibitem

\bibitem{Laine:2007qy}
\begin{barticle}
\bauthor{\bsnm{Laine}, \binits{M.}},
\bauthor{\bsnm{Philipsen}, \binits{O.}},
\bauthor{\bsnm{Tassler}, \binits{M.}}:
\batitle{{Thermal imaginary part of a real-time static potential from classical
  lattice gauge theory simulations}}.
\bjtitle{JHEP}
\bvolume{09},
\bfpage{066}
(\byear{2007}).
doi:\doiurl{10.1088/1126-6708/2007/09/066}.
\arxivurl{0707.2458}
\end{barticle}
\endbibitem

\bibitem{phdthesis_AlexanderLehmann}
\begin{botherref}
\oauthor{\bsnm{Lehmann}, \binits{A.}}:
Minkowskian lattice simulation for non-relativistic quarks in classical fields.
PhD thesis,
Heidelberg University,
Heidelberg
(June 2020).
doi:\doiurl{10.11588/heidok.00028377}
\end{botherref}
\endbibitem

\bibitem{Arnold_1997}
\begin{barticle}
\bauthor{\bsnm{Arnold}, \binits{P.}}:
\batitle{Hotbviolation, the lattice, and hard thermal loops}.
\bjtitle{Physical Review D}
\bvolume{55}(\bissue{12}),
\bfpage{7781}--\blpage{7796}
(\byear{1997}).
doi:\doiurl{10.1103/physrevd.55.7781}
\end{barticle}
\endbibitem

\bibitem{WATSON_1939}
\begin{barticle}
\bauthor{\bsnm{WATSON}, \binits{G.N.}}:
\batitle{Three triple integrals}.
\bjtitle{The Quarterly Journal of Mathematics}
\bvolume{os-10}(\bissue{1}),
\bfpage{266}--\blpage{276}
(\byear{1939}).
doi:\doiurl{10.1093/qmath/os-10.1.266}
\end{barticle}
\endbibitem

\bibitem{Glasser_2000}
\begin{barticle}
\bauthor{\bsnm{Glasser}, \binits{M.L.}},
\bauthor{\bsnm{Boersma}, \binits{J.}}:
\batitle{Exact values for the cubic lattice green functions}.
\bjtitle{Journal of Physics A: Mathematical and General}
\bvolume{33}(\bissue{28}),
\bfpage{5017}--\blpage{5023}
(\byear{2000}).
doi:\doiurl{10.1088/0305-4470/33/28/306}
\end{barticle}
\endbibitem

\bibitem{Bodeker:1998hm}
\begin{barticle}
\bauthor{\bsnm{Bodeker}, \binits{D.}}:
\batitle{{On the effective dynamics of soft nonAbelian gauge fields at finite
  temperature}}.
\bjtitle{Phys. Lett. B}
\bvolume{426},
\bfpage{351}--\blpage{360}
(\byear{1998}).
doi:\doiurl{10.1016/S0370-2693(98)00279-2}.
\arxivurl{hep-ph/9801430}
\end{barticle}
\endbibitem

\bibitem{Bodeker:1999gx}
\begin{barticle}
\bauthor{\bsnm{Bodeker}, \binits{D.}},
\bauthor{\bsnm{Moore}, \binits{G.D.}},
\bauthor{\bsnm{Rummukainen}, \binits{K.}}:
\batitle{{Chern-Simons number diffusion and hard thermal loops on the
  lattice}}.
\bjtitle{Phys. Rev. D}
\bvolume{61},
\bfpage{056003}
(\byear{2000}).
doi:\doiurl{10.1103/PhysRevD.61.056003}.
\arxivurl{hep-ph/9907545}
\end{barticle}
\endbibitem

\bibitem{huckel1923theory}
\begin{barticle}
\bauthor{\bsnm{H{\"u}ckel}, \binits{E.}},
\bauthor{\bsnm{Debye}, \binits{P.}}:
\batitle{The theory of electrolytes: I. lowering of freezing point and related
  phenomena}.
\bjtitle{Phys. Z}
\bvolume{24},
\bfpage{185}--\blpage{206}
(\byear{1923})
\end{barticle}
\endbibitem

\bibitem{kasper_fermion_2014}
\begin{barticle}
\bauthor{\bsnm{Kasper}, \binits{V.}},
\bauthor{\bsnm{Hebenstreit}, \binits{F.}},
\bauthor{\bsnm{Berges}, \binits{J.}}:
\batitle{Fermion production from real-time lattice gauge theory in the
  classical-statistical regime}.
\bjtitle{Physical Review D}
\bvolume{90}(\bissue{2}),
\bfpage{025016}
(\byear{2014}).
doi:\doiurl{10.1103/PhysRevD.90.025016}.
\bcomment{arXiv: 1403.4849}.
Accessed 2020-12-06
\end{barticle}
\endbibitem

\bibitem{Wilson_1974}
\begin{barticle}
\bauthor{\bsnm{Wilson}, \binits{K.G.}}:
\batitle{Confinement of quarks}.
\bjtitle{Physical Review D}
\bvolume{10}(\bissue{8}),
\bfpage{2445}--\blpage{2459}
(\byear{1974}).
doi:\doiurl{10.1103/physrevd.10.2445}
\end{barticle}
\endbibitem

\bibitem{Klassen_1998}
\begin{barticle}
\bauthor{\bsnm{Klassen}, \binits{T.R.}}:
\batitle{The anisotropic wilson gauge action}.
\bjtitle{Nuclear Physics B}
\bvolume{533}(\bissue{1-3}),
\bfpage{557}--\blpage{575}
(\byear{1998}).
doi:\doiurl{10.1016/s0550-3213(98)00510-0}
\end{barticle}
\endbibitem

\bibitem{Ambjorn:1990pu}
\begin{barticle}
\bauthor{\bsnm{Ambjorn}, \binits{J.}},
\bauthor{\bsnm{Askgaard}, \binits{T.}},
\bauthor{\bsnm{Porter}, \binits{H.}},
\bauthor{\bsnm{Shaposhnikov}, \binits{M.E.}}:
\batitle{{Sphaleron transitions and baryon asymmetry: A Numerical real time
  analysis}}.
\bjtitle{Nucl. Phys. B}
\bvolume{353},
\bfpage{346}--\blpage{378}
(\byear{1991}).
doi:\doiurl{10.1016/0550-3213(91)90341-T}
\end{barticle}
\endbibitem

\bibitem{Ambjorn:1997jz}
\begin{barticle}
\bauthor{\bsnm{Ambjorn}, \binits{J.}},
\bauthor{\bsnm{Krasnitz}, \binits{A.}}:
\batitle{{Improved determination of the classical sphaleron transition rate}}.
\bjtitle{Nucl. Phys. B}
\bvolume{506},
\bfpage{387}--\blpage{403}
(\byear{1997}).
doi:\doiurl{10.1016/S0550-3213(97)00524-5}.
\arxivurl{hep-ph/9705380}
\end{barticle}
\endbibitem

\bibitem{Grigoriev:1988bd}
\begin{barticle}
\bauthor{\bsnm{Grigoriev}, \binits{D.Y.}},
\bauthor{\bsnm{Rubakov}, \binits{V.A.}}:
\batitle{{Soliton Pair Creation at Finite Temperatures. Numerical Study in
  (1+1)-dimensions}}.
\bjtitle{Nucl. Phys. B}
\bvolume{299},
\bfpage{67}--\blpage{78}
(\byear{1988}).
doi:\doiurl{10.1016/0550-3213(88)90466-X}
\end{barticle}
\endbibitem

\bibitem{Akamatsu:2015kau}
\begin{barticle}
\bauthor{\bsnm{Akamatsu}, \binits{Y.}},
\bauthor{\bsnm{Rothkopf}, \binits{A.}},
\bauthor{\bsnm{Yamamoto}, \binits{N.}}:
\batitle{{Non-Abelian chiral instabilities at high temperature on the
  lattice}}.
\bjtitle{JHEP}
\bvolume{03},
\bfpage{210}
(\byear{2016}).
doi:\doiurl{10.1007/JHEP03(2016)210}.
\arxivurl{1512.02374}
\end{barticle}
\endbibitem

\bibitem{Hann_window}
\begin{barticle}
\bauthor{\bsnm{{Blackman}}, \binits{R.B.}},
\bauthor{\bsnm{{Tukey}}, \binits{J.W.}}:
\batitle{The measurement of power spectra from the point of view of
  communications engineering — part i}.
\bjtitle{The Bell System Technical Journal}
\bvolume{37}(\bissue{1}),
\bfpage{185}--\blpage{282}
(\byear{1958}).
doi:\doiurl{10.1002/j.1538-7305.1958.tb03874.x}
\end{barticle}
\endbibitem

\bibitem{alexander_rothkopf_2020_4332406}
\begin{botherref}
\oauthor{\bsnm{Rothkopf}, \binits{A.}}:
"Classical Statistical Simulation of SU(3) Yang-Mills Theory in the Presence of
  Static Sources".
doi:\doiurl{10.5281/zenodo.4332406}.
\url{https://doi.org/10.5281/zenodo.4332406}
\end{botherref}
\endbibitem

\bibitem{yanagihara_distribution_2019}
\begin{barticle}
\bauthor{\bsnm{Yanagihara}, \binits{R.}},
\bauthor{\bsnm{Iritani}, \binits{T.}},
\bauthor{\bsnm{Kitazawa}, \binits{M.}},
\bauthor{\bsnm{Asakawa}, \binits{M.}},
\bauthor{\bsnm{Hatsuda}, \binits{T.}}:
\batitle{Distribution of {Stress} {Tensor} around {Static}
  {Quark}--{Anti}-{Quark} from {Yang}-{Mills} {Gradient} {Flow}}.
\bjtitle{Physics Letters B}
\bvolume{789},
\bfpage{210}--\blpage{214}
(\byear{2019}).
doi:\doiurl{10.1016/j.physletb.2018.09.067}.
\bcomment{arXiv: 1803.05656}.
Accessed 2020-03-21
\end{barticle}
\endbibitem

\bibitem{Lehmann:2020kjg}
\begin{barticle}
\bauthor{\bsnm{Lehmann}, \binits{A.}},
\bauthor{\bsnm{Rothkopf}, \binits{A.}}:
\batitle{{Real-Time-Evolution of Heavy-Quarkonium Bound States}}.
\bjtitle{PoS}
\bvolume{LATTICE2019},
\bfpage{074}
(\byear{2019}).
doi:\doiurl{10.22323/1.363.0074}.
\arxivurl{2003.02509}
\end{barticle}
\endbibitem

\end{thebibliography}

\newcommand{\BMCxmlcomment}[1]{}

\BMCxmlcomment{

<refgrp>

<bibl id="B1">
  <title><p>{Heavy Quarkonium: Progress, Puzzles, and
  Opportunities}</p></title>
  <aug>
    <au><snm>Brambilla</snm><fnm>N.</fnm></au>
    <au><cnm>others</cnm></au>
  </aug>
  <source>Eur. Phys. J.</source>
  <pubdate>2011</pubdate>
  <volume>C71</volume>
  <fpage>1534</fpage>
</bibl>

<bibl id="B2">
  <title><p>{Heavy Quarkonium in Extreme Conditions}</p></title>
  <aug>
    <au><snm>Rothkopf</snm><fnm>A</fnm></au>
  </aug>
  <source>Phys. Rept.</source>
  <pubdate>2020</pubdate>
  <volume>858</volume>
  <fpage>1</fpage>
  <lpage>-117</lpage>
</bibl>

<bibl id="B3">
  <title><p>{Pseudothermalization of the quark-gluon plasma}</p></title>
  <aug>
    <au><cnm>{Strickland, Michael}</cnm></au>
  </aug>
  <source>J. Phys. Conf. Ser.</source>
  <editor>{Bellwied, Rene and Harris, John and Ratti, Claudia and Timmins,
  Anthony}</editor>
  <pubdate>2020</pubdate>
  <volume>1602</volume>
  <issue>1</issue>
  <fpage>012018</fpage>
</bibl>

<bibl id="B4">
  <title><p>{Thermalization in QCD: theoretical approaches, phenomenological
  applications, and interdisciplinary connections}</p></title>
  <aug>
    <au><snm>Berges</snm><fnm>J</fnm></au>
    <au><snm>Heller</snm><fnm>MP</fnm></au>
    <au><snm>Mazeliauskas</snm><fnm>A</fnm></au>
    <au><snm>Venugopalan</snm><fnm>R</fnm></au>
  </aug>
  <pubdate>2020</pubdate>
</bibl>

<bibl id="B5">
  <title><p>{Studying QGP with flow: A theory overview}</p></title>
  <aug>
    <au><snm>Shen</snm><fnm>C</fnm></au>
  </aug>
  <source>28th International Conference on Ultrarelativistic Nucleus-Nucleus
  Collisions</source>
  <pubdate>2020</pubdate>
</bibl>

<bibl id="B6">
  <title><p>{Approach to thermalization and hydrodynamics}</p></title>
  <aug>
    <au><snm>Akamatsu</snm><fnm>Y</fnm></au>
  </aug>
  <source>28th International Conference on Ultrarelativistic Nucleus-Nucleus
  Collisions</source>
  <pubdate>2020</pubdate>
</bibl>

<bibl id="B7">
  <title><p>{Matching the non-equilibrium initial stage of heavy ion collisions
  to hydrodynamics with QCD kinetic theory}</p></title>
  <aug>
    <au><snm>Kurkela</snm><fnm>A</fnm></au>
    <au><snm>Mazeliauskas</snm><fnm>A</fnm></au>
    <au><snm>Paquet</snm><fnm>JF</fnm></au>
    <au><snm>Schlichting</snm><fnm>S</fnm></au>
    <au><snm>Teaney</snm><fnm>D</fnm></au>
  </aug>
  <source>PoS</source>
  <pubdate>2019</pubdate>
  <volume>Confinement2018</volume>
  <fpage>152</fpage>
</bibl>

<bibl id="B8">
  <title><p>{Excited bottomonia in quark-gluon plasma from lattice
  QCD}</p></title>
  <aug>
    <au><snm>Larsen</snm><fnm>R</fnm></au>
    <au><snm>Meinel</snm><fnm>S</fnm></au>
    <au><snm>Mukherjee</snm><fnm>S</fnm></au>
    <au><snm>Petreczky</snm><fnm>P</fnm></au>
  </aug>
  <source>Phys. Lett. B</source>
  <pubdate>2020</pubdate>
  <volume>800</volume>
  <fpage>135119</fpage>
</bibl>

<bibl id="B9">
  <title><p>{News from bottomonium spectral functions in thermal
  QCD}</p></title>
  <aug>
    <au><snm>Offler</snm><fnm>S</fnm></au>
    <au><snm>Aarts</snm><fnm>G</fnm></au>
    <au><snm>Allton</snm><fnm>C</fnm></au>
    <au><snm>Glesaaen</snm><fnm>J</fnm></au>
    <au><snm>J\"ager</snm><fnm>B</fnm></au>
    <au><snm>Kim</snm><fnm>S</fnm></au>
    <au><snm>Lombardo</snm><fnm>MP</fnm></au>
    <au><snm>Ryan</snm><fnm>SM</fnm></au>
    <au><snm>Skullerud</snm><fnm>JI</fnm></au>
  </aug>
  <source>PoS</source>
  <pubdate>2019</pubdate>
  <volume>LATTICE2019</volume>
  <fpage>076</fpage>
</bibl>

<bibl id="B10">
  <title><p>{Thermal broadening of bottomonia: Lattice nonrelativistic QCD with
  extended operators}</p></title>
  <aug>
    <au><snm>Larsen</snm><fnm>R</fnm></au>
    <au><snm>Meinel</snm><fnm>S</fnm></au>
    <au><snm>Mukherjee</snm><fnm>S</fnm></au>
    <au><snm>Petreczky</snm><fnm>P</fnm></au>
  </aug>
  <source>Phys. Rev. D</source>
  <pubdate>2019</pubdate>
  <volume>100</volume>
  <issue>7</issue>
  <fpage>074506</fpage>
</bibl>

<bibl id="B11">
  <title><p>{Quarkonium in-medium properties from realistic lattice
  NRQCD}</p></title>
  <aug>
    <au><snm>Kim</snm><fnm>S</fnm></au>
    <au><snm>Petreczky</snm><fnm>P</fnm></au>
    <au><snm>Rothkopf</snm><fnm>A</fnm></au>
  </aug>
  <source>JHEP</source>
  <pubdate>2018</pubdate>
  <volume>11</volume>
  <fpage>088</fpage>
</bibl>

<bibl id="B12">
  <title><p>{Thermal quarkonium physics in the pseudoscalar
  channel}</p></title>
  <aug>
    <au><snm>Burnier</snm><fnm>Y.</fnm></au>
    <au><snm>Ding</snm><fnm>H. T.</fnm></au>
    <au><snm>Kaczmarek</snm><fnm>O.</fnm></au>
    <au><snm>Kruse</snm><fnm>A. L.</fnm></au>
    <au><snm>Laine</snm><fnm>M.</fnm></au>
    <au><snm>Ohno</snm><fnm>H.</fnm></au>
    <au><snm>Sandmeyer</snm><fnm>H.</fnm></au>
  </aug>
  <source>JHEP</source>
  <pubdate>2017</pubdate>
  <volume>11</volume>
  <fpage>206</fpage>
</bibl>

<bibl id="B13">
  <title><p>{Quarkonium at finite temperature: Towards realistic phenomenology
  from first principles}</p></title>
  <aug>
    <au><snm>Burnier</snm><fnm>Y</fnm></au>
    <au><snm>Kaczmarek</snm><fnm>O</fnm></au>
    <au><snm>Rothkopf</snm><fnm>A</fnm></au>
  </aug>
  <source>JHEP</source>
  <pubdate>2015</pubdate>
  <volume>12</volume>
  <fpage>101</fpage>
</bibl>

<bibl id="B14">
  <title><p>{In-medium P-wave quarkonium from the complex lattice QCD
  potential}</p></title>
  <aug>
    <au><snm>Burnier</snm><fnm>Y</fnm></au>
    <au><snm>Kaczmarek</snm><fnm>O</fnm></au>
    <au><snm>Rothkopf</snm><fnm>A</fnm></au>
  </aug>
  <source>JHEP</source>
  <pubdate>2016</pubdate>
  <volume>10</volume>
  <fpage>032</fpage>
</bibl>

<bibl id="B15">
  <title><p>{$J/\psi$ near $T_c$}</p></title>
  <aug>
    <au><snm>Song</snm><fnm>T</fnm></au>
    <au><snm>Gubler</snm><fnm>P</fnm></au>
    <au><snm>Hong</snm><fnm>J</fnm></au>
    <au><snm>Lee</snm><fnm>SH</fnm></au>
    <au><snm>Morita</snm><fnm>K</fnm></au>
  </aug>
  <pubdate>2020</pubdate>
</bibl>

<bibl id="B16">
  <title><p>{Holographic model for charmonium dissociation}</p></title>
  <aug>
    <au><snm>Braga</snm><fnm>NRF</fnm></au>
    <au><snm>Ferreira</snm><fnm>LF</fnm></au>
    <au><snm>Vega</snm><fnm>A</fnm></au>
  </aug>
  <source>Phys. Lett.</source>
  <pubdate>2017</pubdate>
  <volume>B774</volume>
  <fpage>476</fpage>
  <lpage>481</lpage>
</bibl>

<bibl id="B17">
  <title><p>{Towards the Gravity Dual of Quarkonium in the Strongly Coupled QCD
  Plasma}</p></title>
  <aug>
    <au><snm>Grigoryan</snm><fnm>HR</fnm></au>
    <au><snm>Hohler</snm><fnm>PM</fnm></au>
    <au><snm>Stephanov</snm><fnm>MA</fnm></au>
  </aug>
  <source>Phys. Rev.</source>
  <pubdate>2010</pubdate>
  <volume>D82</volume>
  <fpage>026005</fpage>
</bibl>

<bibl id="B18">
  <title><p>{Melting Spectral Functions of the Scalar and Vector Mesons in a
  Holographic QCD Model}</p></title>
  <aug>
    <au><snm>Fujita</snm><fnm>M</fnm></au>
    <au><snm>Kikuchi</snm><fnm>T</fnm></au>
    <au><snm>Fukushima</snm><fnm>K</fnm></au>
    <au><snm>Misumi</snm><fnm>T</fnm></au>
    <au><snm>Murata</snm><fnm>M</fnm></au>
  </aug>
  <source>Phys. Rev.</source>
  <pubdate>2010</pubdate>
  <volume>D81</volume>
  <fpage>065024</fpage>
</bibl>

<bibl id="B19">
  <title><p>{Heavy quarkonia in a medium as a quantum dissipative system:
  Master equation approach}</p></title>
  <aug>
    <au><snm>Borghini</snm><fnm>N</fnm></au>
    <au><snm>Gombeaud</snm><fnm>C</fnm></au>
  </aug>
  <source>Eur. Phys. J.</source>
  <pubdate>2012</pubdate>
  <volume>C72</volume>
  <fpage>2000</fpage>
</bibl>

<bibl id="B20">
  <title><p>{Stochastic potential and quantum decoherence of heavy quarkonium
  in the quark-gluon plasma}</p></title>
  <aug>
    <au><snm>Akamatsu</snm><fnm>Y</fnm></au>
    <au><snm>Rothkopf</snm><fnm>A</fnm></au>
  </aug>
  <source>Phys. Rev.</source>
  <pubdate>2012</pubdate>
  <volume>D85</volume>
  <fpage>105011</fpage>
</bibl>

<bibl id="B21">
  <title><p>{Real-time quantum dynamics of heavy quark systems at high
  temperature}</p></title>
  <aug>
    <au><snm>Akamatsu</snm><fnm>Y</fnm></au>
  </aug>
  <source>Phys. Rev.</source>
  <pubdate>2013</pubdate>
  <volume>D87</volume>
  <issue>4</issue>
  <fpage>045016</fpage>
</bibl>

<bibl id="B22">
  <title><p>{Bottomonium suppression in an open quantum system using the
  quantum trajectories method}</p></title>
  <aug>
    <au><snm>Brambilla</snm><fnm>N</fnm></au>
    <au><snm>Escobedo</snm><fnm>MA</fnm></au>
    <au><snm>Strickland</snm><fnm>M</fnm></au>
    <au><snm>Vairo</snm><fnm>A</fnm></au>
    <au><snm>Vander Griend</snm><fnm>P</fnm></au>
    <au><snm>Weber</snm><fnm>JH</fnm></au>
  </aug>
  <pubdate>2020</pubdate>
</bibl>

<bibl id="B23">
  <title><p>{Quantum simulation of open quantum systems in heavy-ion
  collisions}</p></title>
  <aug>
    <au><snm>De Jong</snm><fnm>WA</fnm></au>
    <au><snm>Metcalf</snm><fnm>M</fnm></au>
    <au><snm>Mulligan</snm><fnm>J</fnm></au>
    <au><snm>P\l{}osko\'n</snm><fnm>M</fnm></au>
    <au><snm>Ringer</snm><fnm>F</fnm></au>
    <au><snm>Yao</snm><fnm>X</fnm></au>
  </aug>
  <pubdate>2020</pubdate>
</bibl>

<bibl id="B24">
  <title><p>{Quarkonium Semiclassical Transport in Quark-Gluon Plasma:
  Factorization and Quantum Correction}</p></title>
  <aug>
    <au><snm>Yao</snm><fnm>X</fnm></au>
    <au><snm>Mehen</snm><fnm>T</fnm></au>
  </aug>
  <pubdate>2020</pubdate>
</bibl>

<bibl id="B25">
  <title><p>{Coupled Boltzmann Transport Equations of Heavy Quarks and
  Quarkonia in Quark-Gluon Plasma}</p></title>
  <aug>
    <au><snm>Yao</snm><fnm>X</fnm></au>
    <au><snm>Ke</snm><fnm>W</fnm></au>
    <au><snm>Xu</snm><fnm>Y</fnm></au>
    <au><snm>Bass</snm><fnm>SA</fnm></au>
    <au><snm>M\"uller</snm><fnm>B</fnm></au>
  </aug>
  <pubdate>2020</pubdate>
</bibl>

<bibl id="B26">
  <title><p>{Trace preserving quantum dynamics using a novel
  reparametrization-neutral summation-by-parts difference operator}</p></title>
  <aug>
    <au><snm>\r{A}lund</snm><fnm>O</fnm></au>
    <au><snm>Akamatsu</snm><fnm>Y</fnm></au>
    <au><snm>Laur\'en</snm><fnm>F</fnm></au>
    <au><snm>Miura</snm><fnm>T</fnm></au>
    <au><snm>Nordstr\"om</snm><fnm>J</fnm></au>
    <au><snm>Rothkopf</snm><fnm>A</fnm></au>
  </aug>
  <pubdate>2020</pubdate>
</bibl>

<bibl id="B27">
  <title><p>{Quantum Brownian motion of a heavy quark pair in the quark-gluon
  plasma}</p></title>
  <aug>
    <au><snm>Miura</snm><fnm>T</fnm></au>
    <au><snm>Akamatsu</snm><fnm>Y</fnm></au>
    <au><snm>Asakawa</snm><fnm>M</fnm></au>
    <au><snm>Rothkopf</snm><fnm>A</fnm></au>
  </aug>
  <source>Phys. Rev. D</source>
  <pubdate>2020</pubdate>
  <volume>101</volume>
  <issue>3</issue>
  <fpage>034011</fpage>
</bibl>

<bibl id="B28">
  <title><p>{Transport coefficients from in medium quarkonium
  dynamics}</p></title>
  <aug>
    <au><snm>Brambilla</snm><fnm>N</fnm></au>
    <au><snm>Escobedo</snm><fnm>MA</fnm></au>
    <au><snm>Vairo</snm><fnm>A</fnm></au>
    <au><snm>Vander Griend</snm><fnm>P</fnm></au>
  </aug>
  <source>Phys. Rev. D</source>
  <pubdate>2019</pubdate>
  <volume>100</volume>
  <issue>5</issue>
  <fpage>054025</fpage>
</bibl>

<bibl id="B29">
  <title><p>{Quarkonium inside the quark-gluon plasma: Diffusion, dissociation,
  recombination, and energy loss}</p></title>
  <aug>
    <au><snm>Yao</snm><fnm>X</fnm></au>
    <au><snm>Müller</snm><fnm>B</fnm></au>
  </aug>
  <source>Phys. Rev.</source>
  <pubdate>2019</pubdate>
  <volume>D100</volume>
  <issue>1</issue>
  <fpage>014008</fpage>
</bibl>

<bibl id="B30">
  <title><p>{Approach to equilibrium of a quarkonium in a quark-gluon
  plasma}</p></title>
  <aug>
    <au><snm>Blaizot</snm><fnm>JP</fnm></au>
    <au><snm>Escobedo</snm><fnm>MA</fnm></au>
  </aug>
  <source>Phys. Rev.</source>
  <pubdate>2018</pubdate>
  <volume>D98</volume>
  <issue>7</issue>
  <fpage>074007</fpage>
</bibl>

<bibl id="B31">
  <title><p>{Heavy quarkonium suppression in a fireball}</p></title>
  <aug>
    <au><snm>Brambilla</snm><fnm>N</fnm></au>
    <au><snm>Escobedo</snm><fnm>MA</fnm></au>
    <au><snm>Soto</snm><fnm>J</fnm></au>
    <au><snm>Vairo</snm><fnm>A</fnm></au>
  </aug>
  <source>Phys. Rev. D</source>
  <pubdate>2018</pubdate>
  <volume>97</volume>
  <issue>7</issue>
  <fpage>074009</fpage>
</bibl>

<bibl id="B32">
  <title><p>{Dynamical dissociation of quarkonia by wave function
  decoherence}</p></title>
  <aug>
    <au><snm>Kajimoto</snm><fnm>S</fnm></au>
    <au><snm>Akamatsu</snm><fnm>Y</fnm></au>
    <au><snm>Asakawa</snm><fnm>M</fnm></au>
    <au><snm>Rothkopf</snm><fnm>A</fnm></au>
  </aug>
  <source>Phys. Rev. D</source>
  <pubdate>2018</pubdate>
  <volume>97</volume>
  <issue>1</issue>
  <fpage>014003</fpage>
</bibl>

<bibl id="B33">
  <title><p>{Approach to equilibrium of quarkonium in quark-gluon
  plasma}</p></title>
  <aug>
    <au><snm>Yao</snm><fnm>X</fnm></au>
    <au><snm>Müller</snm><fnm>B</fnm></au>
  </aug>
  <source>Phys. Rev.</source>
  <pubdate>2018</pubdate>
  <volume>C97</volume>
  <issue>1</issue>
  <fpage>014908</fpage>
  <note>[Erratum: Phys. Rev.C97,no.4,049903(2018)]</note>
</bibl>

<bibl id="B34">
  <title><p>{Quarkonium in Quark-Gluon Plasma: Open Quantum System Approaches
  Re-examined}</p></title>
  <aug>
    <au><snm>Akamatsu</snm><fnm>Y</fnm></au>
  </aug>
  <pubdate>2020</pubdate>
</bibl>

<bibl id="B35">
  <title><p>{Effective field theories for heavy quarkonium}</p></title>
  <aug>
    <au><snm>Brambilla</snm><fnm>N</fnm></au>
    <au><snm>Pineda</snm><fnm>A</fnm></au>
    <au><snm>Soto</snm><fnm>J</fnm></au>
    <au><snm>Vairo</snm><fnm>A</fnm></au>
  </aug>
  <source>Rev. Mod. Phys.</source>
  <pubdate>2005</pubdate>
  <volume>77</volume>
  <fpage>1423</fpage>
</bibl>

<bibl id="B36">
  <title><p>{Real-time static potential in hot QCD}</p></title>
  <aug>
    <au><snm>Laine</snm><fnm>M.</fnm></au>
    <au><snm>Philipsen</snm><fnm>O.</fnm></au>
    <au><snm>Romatschke</snm><fnm>P.</fnm></au>
    <au><snm>Tassler</snm><fnm>M.</fnm></au>
  </aug>
  <source>JHEP</source>
  <pubdate>2007</pubdate>
  <volume>03</volume>
  <fpage>054</fpage>
</bibl>

<bibl id="B37">
  <title><p>{Real and imaginary-time Q anti-Q correlators in a thermal
  medium}</p></title>
  <aug>
    <au><snm>Beraudo</snm><fnm>A.</fnm></au>
    <au><snm>Blaizot</snm><fnm>J. P.</fnm></au>
    <au><snm>Ratti</snm><fnm>C.</fnm></au>
  </aug>
  <source>Nucl. Phys.</source>
  <pubdate>2008</pubdate>
  <volume>A806</volume>
  <fpage>312</fpage>
  <lpage>338</lpage>
</bibl>

<bibl id="B38">
  <title><p>{Static quark-antiquark pairs at finite temperature}</p></title>
  <aug>
    <au><snm>Brambilla</snm><fnm>N</fnm></au>
    <au><snm>Ghiglieri</snm><fnm>J</fnm></au>
    <au><snm>Vairo</snm><fnm>A</fnm></au>
    <au><snm>Petreczky</snm><fnm>P</fnm></au>
  </aug>
  <source>Phys. Rev.</source>
  <pubdate>2008</pubdate>
  <volume>D78</volume>
  <fpage>014017</fpage>
</bibl>

<bibl id="B39">
  <title><p>{Complex Heavy-Quark Potential at Finite Temperature from Lattice
  QCD}</p></title>
  <aug>
    <au><snm>Rothkopf</snm><fnm>A</fnm></au>
    <au><snm>Hatsuda</snm><fnm>T</fnm></au>
    <au><snm>Sasaki</snm><fnm>S</fnm></au>
  </aug>
  <source>Phys. Rev. Lett.</source>
  <pubdate>2012</pubdate>
  <volume>108</volume>
  <fpage>162001</fpage>
</bibl>

<bibl id="B40">
  <title><p>{Static quark-antiquark potential in the quark-gluon plasma from
  lattice QCD}</p></title>
  <aug>
    <au><snm>Burnier</snm><fnm>Y</fnm></au>
    <au><snm>Kaczmarek</snm><fnm>O</fnm></au>
    <au><snm>Rothkopf</snm><fnm>A</fnm></au>
  </aug>
  <source>Phys. Rev. Lett.</source>
  <pubdate>2015</pubdate>
  <volume>114</volume>
  <issue>8</issue>
  <fpage>082001</fpage>
</bibl>

<bibl id="B41">
  <title><p>{Complex heavy-quark potential and Debye mass in a gluonic medium
  from lattice QCD}</p></title>
  <aug>
    <au><snm>Burnier</snm><fnm>Y</fnm></au>
    <au><snm>Rothkopf</snm><fnm>A</fnm></au>
  </aug>
  <source>Phys. Rev. D</source>
  <pubdate>2017</pubdate>
  <volume>95</volume>
  <issue>5</issue>
  <fpage>054511</fpage>
</bibl>

<bibl id="B42">
  <title><p>{Lattice Calculations of Heavy Quark Potential at Finite
  Temperature}</p></title>
  <aug>
    <au><snm>Petreczky</snm><fnm>P.</fnm></au>
    <au><snm>Weber</snm><fnm>J.</fnm></au>
  </aug>
  <source>Nucl. Phys. A</source>
  <editor>Heinz, Ulrich and Evdokimov, Olga and Jacobs, Peter</editor>
  <pubdate>2017</pubdate>
  <volume>967</volume>
  <fpage>592</fpage>
  <lpage>-595</lpage>
</bibl>

<bibl id="B43">
  <title><p>{Realistic in-medium heavy-quark potential from high statistics
  lattice QCD simulations}</p></title>
  <aug>
    <au><snm>Petreczky</snm><fnm>P</fnm></au>
    <au><snm>Rothkopf</snm><fnm>A</fnm></au>
    <au><snm>Weber</snm><fnm>J</fnm></au>
  </aug>
  <source>Nucl. Phys. A</source>
  <editor>Antinori, Federico and Dainese, Andrea and Giubellino, Paolo and
  Greco, Vincenzo and Lombardo, Maria Paola and Scomparin, Enrico</editor>
  <pubdate>2019</pubdate>
  <volume>982</volume>
  <fpage>735</fpage>
  <lpage>-738</lpage>
</bibl>

<bibl id="B44">
  <title><p>{Improved Gauss law model and in-medium heavy quarkonium at finite
  density and velocity}</p></title>
  <aug>
    <au><snm>Lafferty</snm><fnm>D</fnm></au>
    <au><snm>Rothkopf</snm><fnm>A</fnm></au>
  </aug>
  <source>Phys. Rev. D</source>
  <pubdate>2020</pubdate>
  <volume>101</volume>
  <issue>5</issue>
  <fpage>056010</fpage>
</bibl>

<bibl id="B45">
  <title><p>{Disentangling the timescales behind the non-perturbative heavy
  quark potential}</p></title>
  <aug>
    <au><snm>Burnier</snm><fnm>Y</fnm></au>
    <au><snm>Rothkopf</snm><fnm>A</fnm></au>
  </aug>
  <source>Phys. Rev. D</source>
  <pubdate>2012</pubdate>
  <volume>86</volume>
  <fpage>051503</fpage>
</bibl>

<bibl id="B46">
  <title><p>{A hard thermal loop benchmark for the extraction of the
  nonperturbative $Q\bar{Q}$ potential}</p></title>
  <aug>
    <au><snm>Burnier</snm><fnm>Y</fnm></au>
    <au><snm>Rothkopf</snm><fnm>A</fnm></au>
  </aug>
  <source>Phys. Rev. D</source>
  <pubdate>2013</pubdate>
  <volume>87</volume>
  <fpage>114019</fpage>
</bibl>

<bibl id="B47">
  <title><p>{Bayesian techniques and applications to QCD}</p></title>
  <aug>
    <au><snm>Rothkopf</snm><fnm>A</fnm></au>
  </aug>
  <source>PoS</source>
  <pubdate>2018</pubdate>
  <volume>Confinement2018</volume>
  <fpage>026</fpage>
</bibl>

<bibl id="B48">
  <title><p>{Sphaleron Rate in the Minimal Standard Model}</p></title>
  <aug>
    <au><snm>D'Onofrio</snm><fnm>M</fnm></au>
    <au><snm>Rummukainen</snm><fnm>K</fnm></au>
    <au><snm>Tranberg</snm><fnm>A</fnm></au>
  </aug>
  <source>Phys. Rev. Lett.</source>
  <pubdate>2014</pubdate>
  <volume>113</volume>
  <issue>14</issue>
  <fpage>141602</fpage>
</bibl>

<bibl id="B49">
  <title><p>{Turbulent thermalization}</p></title>
  <aug>
    <au><snm>Micha</snm><fnm>R</fnm></au>
    <au><snm>Tkachev</snm><fnm>II</fnm></au>
  </aug>
  <source>Phys. Rev. D</source>
  <pubdate>2004</pubdate>
  <volume>70</volume>
  <fpage>043538</fpage>
</bibl>

<bibl id="B50">
  <title><p>{Non-thermal fixed points: Effective weak-coupling for strongly
  correlated systems far from equilibrium}</p></title>
  <aug>
    <au><snm>Berges</snm><fnm>J</fnm></au>
    <au><snm>Rothkopf</snm><fnm>A</fnm></au>
    <au><snm>Schmidt</snm><fnm>J</fnm></au>
  </aug>
  <source>Phys. Rev. Lett.</source>
  <pubdate>2008</pubdate>
  <volume>101</volume>
  <fpage>041603</fpage>
</bibl>

<bibl id="B51">
  <title><p>{Nonthermal Fixed Points in Quantum Field Theory Beyond the
  Weak-Coupling Limit}</p></title>
  <aug>
    <au><snm>Berges</snm><fnm>J</fnm></au>
    <au><snm>Wallisch</snm><fnm>B</fnm></au>
  </aug>
  <source>Phys. Rev. D</source>
  <pubdate>2017</pubdate>
  <volume>95</volume>
  <issue>3</issue>
  <fpage>036016</fpage>
</bibl>

<bibl id="B52">
  <title><p>{Analog reheating of the early universe in the
  laboratory}</p></title>
  <aug>
    <au><snm>Chatrchyan</snm><fnm>A</fnm></au>
    <au><snm>Geier</snm><fnm>KT</fnm></au>
    <au><snm>Oberthaler</snm><fnm>MK</fnm></au>
    <au><snm>Berges</snm><fnm>J</fnm></au>
    <au><snm>Hauke</snm><fnm>P</fnm></au>
  </aug>
  <pubdate>2020</pubdate>
</bibl>

<bibl id="B53">
  <title><p>{Thermal imaginary part of a real-time static potential from
  classical lattice gauge theory simulations}</p></title>
  <aug>
    <au><snm>Laine</snm><fnm>M.</fnm></au>
    <au><snm>Philipsen</snm><fnm>O.</fnm></au>
    <au><snm>Tassler</snm><fnm>M.</fnm></au>
  </aug>
  <source>JHEP</source>
  <pubdate>2007</pubdate>
  <volume>09</volume>
  <fpage>066</fpage>
</bibl>

<bibl id="B54">
  <title><p>Minkowskian Lattice Simulation for Non-Relativistic Quarks in
  Classical Fields</p></title>
  <aug>
    <au><snm>Lehmann</snm><fnm>A</fnm></au>
  </aug>
  <source>PhD thesis</source>
  <publisher>Heidelberg University</publisher>
  <pubdate>2020</pubdate>
</bibl>

<bibl id="B55">
  <title><p>HotBviolation, the lattice, and hard thermal loops</p></title>
  <aug>
    <au><snm>Arnold</snm><fnm>P</fnm></au>
  </aug>
  <source>Physical Review D</source>
  <publisher>American Physical Society (APS)</publisher>
  <pubdate>1997</pubdate>
  <volume>55</volume>
  <issue>12</issue>
  <fpage>7781–7796</fpage>
  <url>http://dx.doi.org/10.1103/PhysRevD.55.7781</url>
</bibl>

<bibl id="B56">
  <title><p>THREE TRIPLE INTEGRALS</p></title>
  <aug>
    <au><snm>WATSON</snm><fnm>G. N.</fnm></au>
  </aug>
  <source>The Quarterly Journal of Mathematics</source>
  <publisher>Oxford University Press (OUP)</publisher>
  <pubdate>1939</pubdate>
  <volume>os-10</volume>
  <issue>1</issue>
  <fpage>266–276</fpage>
  <url>http://dx.doi.org/10.1093/QMATH/OS-10.1.266</url>
</bibl>

<bibl id="B57">
  <title><p>Exact values for the cubic lattice Green functions</p></title>
  <aug>
    <au><snm>Glasser</snm><fnm>M L</fnm></au>
    <au><snm>Boersma</snm><fnm>J</fnm></au>
  </aug>
  <source>Journal of Physics A: Mathematical and General</source>
  <publisher>IOP Publishing</publisher>
  <pubdate>2000</pubdate>
  <volume>33</volume>
  <issue>28</issue>
  <fpage>5017–5023</fpage>
  <url>http://dx.doi.org/10.1088/0305-4470/33/28/306</url>
</bibl>

<bibl id="B58">
  <title><p>{On the effective dynamics of soft nonAbelian gauge fields at
  finite temperature}</p></title>
  <aug>
    <au><snm>Bodeker</snm><fnm>D</fnm></au>
  </aug>
  <source>Phys. Lett. B</source>
  <pubdate>1998</pubdate>
  <volume>426</volume>
  <fpage>351</fpage>
  <lpage>-360</lpage>
</bibl>

<bibl id="B59">
  <title><p>{Chern-Simons number diffusion and hard thermal loops on the
  lattice}</p></title>
  <aug>
    <au><snm>Bodeker</snm><fnm>D.</fnm></au>
    <au><snm>Moore</snm><fnm>GD</fnm></au>
    <au><snm>Rummukainen</snm><fnm>K.</fnm></au>
  </aug>
  <source>Phys. Rev. D</source>
  <pubdate>2000</pubdate>
  <volume>61</volume>
  <fpage>056003</fpage>
</bibl>

<bibl id="B60">
  <title><p>The theory of electrolytes: I. lowering of freezing point and
  related phenomena</p></title>
  <aug>
    <au><snm>H{\"u}ckel</snm><fnm>E</fnm></au>
    <au><snm>Debye</snm><fnm>P</fnm></au>
  </aug>
  <source>Phys. Z</source>
  <pubdate>1923</pubdate>
  <volume>24</volume>
  <fpage>185</fpage>
  <lpage>206</lpage>
</bibl>

<bibl id="B61">
  <title><p>Fermion production from real-time lattice gauge theory in the
  classical-statistical regime</p></title>
  <aug>
    <au><snm>Kasper</snm><fnm>V</fnm></au>
    <au><snm>Hebenstreit</snm><fnm>F</fnm></au>
    <au><snm>Berges</snm><fnm>J</fnm></au>
  </aug>
  <source>Physical Review D</source>
  <pubdate>2014</pubdate>
  <volume>90</volume>
  <issue>2</issue>
  <fpage>025016</fpage>
  <url>http://arxiv.org/abs/1403.4849</url>
  <note>arXiv: 1403.4849</note>
</bibl>

<bibl id="B62">
  <title><p>Confinement of quarks</p></title>
  <aug>
    <au><snm>Wilson</snm><fnm>KG</fnm></au>
  </aug>
  <source>Physical Review D</source>
  <publisher>American Physical Society (APS)</publisher>
  <pubdate>1974</pubdate>
  <volume>10</volume>
  <issue>8</issue>
  <fpage>2445–2459</fpage>
  <url>http://dx.doi.org/10.1103/PHYSREVD.10.2445</url>
</bibl>

<bibl id="B63">
  <title><p>The anisotropic Wilson gauge action</p></title>
  <aug>
    <au><snm>Klassen</snm><fnm>TR</fnm></au>
  </aug>
  <source>Nuclear Physics B</source>
  <publisher>Elsevier BV</publisher>
  <pubdate>1998</pubdate>
  <volume>533</volume>
  <issue>1-3</issue>
  <fpage>557–575</fpage>
  <url>http://dx.doi.org/10.1016/S0550-3213(98)00510-0</url>
</bibl>

<bibl id="B64">
  <title><p>{Sphaleron transitions and baryon asymmetry: A Numerical real time
  analysis}</p></title>
  <aug>
    <au><snm>Ambjorn</snm><fnm>J</fnm></au>
    <au><snm>Askgaard</snm><fnm>T.</fnm></au>
    <au><snm>Porter</snm><fnm>H.</fnm></au>
    <au><snm>Shaposhnikov</snm><fnm>M.E.</fnm></au>
  </aug>
  <source>Nucl. Phys. B</source>
  <pubdate>1991</pubdate>
  <volume>353</volume>
  <fpage>346</fpage>
  <lpage>-378</lpage>
</bibl>

<bibl id="B65">
  <title><p>{Improved determination of the classical sphaleron transition
  rate}</p></title>
  <aug>
    <au><snm>Ambjorn</snm><fnm>J</fnm></au>
    <au><snm>Krasnitz</snm><fnm>A.</fnm></au>
  </aug>
  <source>Nucl. Phys. B</source>
  <pubdate>1997</pubdate>
  <volume>506</volume>
  <fpage>387</fpage>
  <lpage>-403</lpage>
</bibl>

<bibl id="B66">
  <title><p>{Soliton Pair Creation at Finite Temperatures. Numerical Study in
  (1+1)-dimensions}</p></title>
  <aug>
    <au><snm>Grigoriev</snm><fnm>DY</fnm></au>
    <au><snm>Rubakov</snm><fnm>V.A.</fnm></au>
  </aug>
  <source>Nucl. Phys. B</source>
  <pubdate>1988</pubdate>
  <volume>299</volume>
  <fpage>67</fpage>
  <lpage>-78</lpage>
</bibl>

<bibl id="B67">
  <title><p>{Non-Abelian chiral instabilities at high temperature on the
  lattice}</p></title>
  <aug>
    <au><snm>Akamatsu</snm><fnm>Y</fnm></au>
    <au><snm>Rothkopf</snm><fnm>A</fnm></au>
    <au><snm>Yamamoto</snm><fnm>N</fnm></au>
  </aug>
  <source>JHEP</source>
  <pubdate>2016</pubdate>
  <volume>03</volume>
  <fpage>210</fpage>
</bibl>

<bibl id="B68">
  <title><p>The measurement of power spectra from the point of view of
  communications engineering — Part I</p></title>
  <aug>
    <au><snm>{Blackman}</snm><fnm>R. B.</fnm></au>
    <au><snm>{Tukey}</snm><fnm>J. W.</fnm></au>
  </aug>
  <source>The Bell System Technical Journal</source>
  <pubdate>1958</pubdate>
  <volume>37</volume>
  <issue>1</issue>
  <fpage>185</fpage>
  <lpage>282</lpage>
</bibl>

<bibl id="B69">
  <title><p>"Classical Statistical Simulation of SU(3) Yang-Mills theory in the
  presence of static sources"</p></title>
  <aug>
    <au><snm>Rothkopf</snm><fnm>A</fnm></au>
  </aug>
  <publisher>Zenodo</publisher>
  <pubdate>2020</pubdate>
  <url>https://doi.org/10.5281/zenodo.4332406</url>
</bibl>

<bibl id="B70">
  <title><p>Distribution of {Stress} {Tensor} around {Static}
  {Quark}--{Anti}-{Quark} from {Yang}-{Mills} {Gradient} {Flow}</p></title>
  <aug>
    <au><snm>Yanagihara</snm><fnm>R</fnm></au>
    <au><snm>Iritani</snm><fnm>T</fnm></au>
    <au><snm>Kitazawa</snm><fnm>M</fnm></au>
    <au><snm>Asakawa</snm><fnm>M</fnm></au>
    <au><snm>Hatsuda</snm><fnm>T</fnm></au>
  </aug>
  <source>Physics Letters B</source>
  <pubdate>2019</pubdate>
  <volume>789</volume>
  <fpage>210</fpage>
  <lpage>-214</lpage>
  <url>http://arxiv.org/abs/1803.05656</url>
  <note>arXiv: 1803.05656</note>
</bibl>

<bibl id="B71">
  <title><p>{Real-Time-Evolution of Heavy-Quarkonium Bound States}</p></title>
  <aug>
    <au><snm>Lehmann</snm><fnm>A</fnm></au>
    <au><snm>Rothkopf</snm><fnm>A</fnm></au>
  </aug>
  <source>PoS</source>
  <pubdate>2019</pubdate>
  <volume>LATTICE2019</volume>
  <fpage>074</fpage>
</bibl>

</refgrp>
} 


\end{backmatter}


\end{document}